\documentclass[12pt,preprint]{aastex}

\begin{document}

\title{Radio Spectral Index Variations and Physical Conditions in Kepler's 
Supernova Remnant}

\author{Tracey DeLaney, Barron Koralesky, and Lawrence Rudnick}
\affil{Department of Astronomy, University of Minnesota, 116 Church Street 
SE,Minneapolis, MN  55455}
\email{tdelaney@astro.umn.edu, barron@astro.umn.edu, larry@astro.umn.edu}
\author{John R. Dickel}
\affil{Astronomy Department, University of Illinois at Urbana-Champaign, 
Urbana, IL  61801}
\email{johnd@astro.uiuc.edu}

\begin{abstract}

A new epoch of VLA measurements of Kepler's supernova remnant was obtained 
to make accurate measurements of the radio spectral index variations and 
polarization.  We have compared these new radio images with H$\alpha$, 
infrared, and X-ray data to better understand the three dimensional structure 
and dynamics of Kepler, and to better understand the physical relationships 
between the various nonthermal and thermal plasmas.  Spatial variations in 
the radio spectral index from $-$0.85 to $-$0.6 are observed between 6 cm and 
20 cm.  The mean spectral index is $-$0.71.  The mean percent polarization 
is 3.5\% at 20 cm and 6\% at 6 cm.  There is a strong correspondence 
between the radial and azimuthal profiles of the radio, X-ray, H$\alpha$, and 
infrared emission in different locations around the remnant although there is 
no single, global pattern.  Spectral tomography shows that the flat- and 
steep-spectrum radio emissions have distinct structures.  The flat-spectrum 
radio emission is found either at a larger radius than or coincident with the 
steep-spectrum emission.  We interpret these spectral components as tracing 
forward- and reverse-shocked material, respectively.  The flat-spectrum radio 
emission can alternatively be interpreted as the bow-shocked material 
(reshocked by the forward shock) from the progenitor's motion through the 
interstellar medium.  The H$\alpha$ and IR images are very similar.  Their 
leading edges are coincident and are either in front of or coincident with 
the leading edges of the X-ray and radio emission.  The X-ray emission 
matches the H$\alpha$ and IR emission in places, and in other places traces 
the steep-spectrum radio emission.  In the north there is also an 
anticorrelation in the azimuthal profiles around the remnant of the 
flat-spectrum radio emission and the thermal X-ray, H$\alpha$, and IR 
emissions.  We suggest that this could be due to a relative weakening of the 
particle acceleration at the forward shock due to Alfv\'{e}n wave damping in 
regions of high density.

\end{abstract}
\keywords{supernova remnants --- ISM: individual (Kepler's SNR) --- radio 
continuum: ISM}


\section{Introduction}
\label{sec:intro}

Kepler's supernova remnant (SNR; G4.5+6.8, SN 1604) is the second youngest 
SNR known in the Galaxy.  Recent distance estimates \citep{rg99} place the 
remnant between 4.8 and 6.4 kpc (we adopt 5 kpc for our calculations) and it 
is about 200$\arcsec$ ($\approx$ 4.8 pc) in diameter.  As a young, bright 
SNR, it has been studied at many wavelengths.  The most recent study 
comparing the radio, X-ray, optical, and infrared (IR) emission was conducted 
by \citet{bra87} at a resolution of 2$\arcmin$.  It is time to revisit this 
analysis using the most recent observations and at a higher resolution.

The radio emission has a shell-like morphology and is brighter in the north 
than in the south. The radio emission results from synchrotron radiation which 
probes the interaction between relativistic particles and the magnetic 
fields.  Shocks and eddy motions at Rayleigh-Taylor instabilities are thought 
to provide the necessary acceleration mechanism and magnetic field 
enhancement to produce the synchrotron emission \citep{gul73a}.  

SNRs are thought to be the primary acceleration sites for Galactic cosmic 
rays below $\approx$ 10$^{14}$ eV.  The acceleration process is probably 
first- or second-order Fermi acceleration in the shocks and/or turbulence 
from the supersonic expansion of the supernova ejecta \citep{jrj98}. Many 
details of the acceleration process are not well established 
observationally.  For instance, we do not know what constitutes the seed 
population of particles for the acceleration mechanism nor how these 
particles are injected into the cycle.  By observing the radio spectral 
index, $\alpha$ ($S_{\nu} \propto \nu^{\alpha}$), of the synchrotron 
radiation from the relativistic electrons, we can determine local shock and 
dynamical characteristics \citep{arl91,ar93,ar96,kkl00}.   

The X-ray emission has the same shell-like morphology and is qualitatively 
similar to the radio emission with the same north-south brightness asymmetry 
\citep{mld84}.  The X-ray emission results from two processes.  The first is 
line emission from heavy elements and the second is bremsstrahlung emission 
from a hot thermal plasma.  At the energies covered by the ROSAT HRI 
($\lesssim$ 2 keV), ASCA observations indicate that Kepler's spectrum 
includes the Fe L line complex and the lines of \ion{Mg}{11} K$\alpha$ and 
\ion{Si}{13} K$\alpha$ \citep{kt99}.  At energies $\lesssim$ 1.4 keV, there 
is also confusion between the unresolved Fe L line complex and the thermal 
continuum.  

Using EINSTEIN and ROSAT HRI images, \citet{hug99} measured the X-ray 
expansion parameter $m$ (defined as $R\propto t^m$) to be 0.93 which 
indicates that the remnant is nearly in free expansion.  The radio expansion 
parameter is $\approx$ 0.5 \citep{dsa88} which is only about half that of the 
X-ray.  The disparity between radio and X-ray expansion appears to be a 
common theme in young SNRs as the same behavior has been found in Cassiopeia 
A \citep{krg98} and Tycho's SNR \citep{hug00}.  

The most recent optical images are over ten years old \citep{blv91, bb91}; 
however, the slow expansion of the optical remnant, $\approx 200$ km s$^{-1}$ 
\citep{bb91}, allows comparison to newer images at radio, X-ray and IR 
wavelengths.  The optical remnant consists of diffuse filaments and bright 
knots.  The emission is concentrated to the north and northwest with some 
structures near the center of the remnant.  Observations indicate both 
radiative and nonradiative (Balmer-dominated) emission \citep{fbb89, blv91}.  
The radiative emission is mostly concentrated in the knotty structures in the 
northwest, where the brightness in all emission lines is strongest, while the 
Balmer-dominated emission is associated with filamentary structures and a few 
knots.  The Balmer-dominated spectrum consists of both broad and narrow 
components.  This type of spectrum results from a high velocity shock passing 
through a low density, partially neutral medium.  The broad component 
specifically results from a charge exchange reaction between slow neutral 
atoms and fast protons \emph{behind} the shock \citep{ckr80}.   
Balmer-dominated emission is referred to as nonradiative because the 
radiative timescale behind the shock is long in comparison with the dynamical 
time scale.  The inferred shock velocity is $\approx$ 1800 km s$^{-1}$.  The 
observed [\ion{S}{2}] line ratios indicate that the optical filaments have 
very high densities ($n_e \ge$ 1000 cm$^{-3}$) compared with other galactic 
SNRs \citep{den82}.

Observations in the IR are important because this emission is thought to be 
the dominant cooling mechanism in SNRs \citep{dra81}.  The most recent IR 
study has been conducted with ISO \citep{dlb99, dlc01}.  The IR continuum 
emission is consistent with a 120 K blackbody.  There are also some weak 
[\ion{Ne}{2}] or [\ion{Ar}{2}] ionic lines in the brightest regions of the 
remnant.  Spatially, the IR emission is strongly correlated with the H$\alpha$ 
emission and is thought to be thermal emission from shock heated dust in the 
circumstellar region.

Kepler's progenitor is somewhat of an enigma.  \citet{baa43} identified the 
remnant as the remains of a type I supernova and its distance above the 
Galactic plane ($\approx$ 600 pc) is consistent with a population II 
progenitor, suggesting a type Ia event.  Also, recent ASCA observations 
\citep{kt99} indicate a relative overabundance of iron that agrees with type 
Ia supernova nucleosynthesis models.  However, the nitrogen overabundance in 
the optical knots \citep{den82}, the slow expansion velocities of these 
knots, and the enhanced density in the region suggest that they are 
circumstellar material ejected by the stellar wind from a massive star.  
\citet{ban87} proposed a model suggesting that the progenitor was a massive 
runaway star.  This model also explains the observed north-south brightness 
asymmetry as a blast wave presently moving through the bow shock (the 
brighter northern rim) in the direction of the progenitor's motion.  
\citet{bbs92} performed numerical simulations which showed that the model 
recovered the observed morphology of Kepler's SNR.

In this paper we wish to provide an improved look at the radio structure, 
spectral variations and polarization properties, as addressed earlier 
by \citet{dsa88} and \citet{mld84}.  We also look in more detail at the 
relationship between the radio emission and those at optical, IR, and X-ray 
wavelengths.  In \S \ref{sec:obs} of the paper we describe the radio 
observations and data reduction.  In \S \ref{sec:radim} we present the various 
images made from the radio data, discuss how they were made, present some 
statistics, and describe features on these images.  The optical, IR, and X-ray 
images are presented in \S \ref{sec:oirx} along with descriptions of features 
on these images.  Comparisons between the various radio images and the 
optical, IR, and X-ray images are presented in \S \ref{sec:comps} with 
discussion in \S \ref{sec:disc} and concluding remarks in \S \ref{sec:conc}.

\section{Radio Observations and Data Reduction}
\label{sec:obs}

Observations were made with the VLA\footnote{The VLA is operated by the 
National Radio Astronomy Observatory, which is a facility of the National 
Science Foundation, operated under cooperative agreement by Associated 
Universities, Inc.} using all four configurations from the most extended 
A-configuration with a maximum antenna separation of 36.4 km to the most 
compact D-configuration with a shortest separation of 35 m.  These are 
summarized in Table \ref{tbl1}.  Data were taken at 4835.1 and 4885.1 MHz 
(hereafter called 6 cm) and at 1285.0, 1365.1, 1402.5, and 1464.9 MHz 
(hereafter called 20 cm).

Standard calibration procedures were used for the 6 cm data and most of the 20 
cm data as described in the AIPS 
Cookbook\footnote{\url{http://www.cv.nrao.edu/aips/cook.html}}.  There were 
two exceptions to this at 20 cm.  First, at A configuration, no flux density 
and polarization data for the primary calibrator were available with the same 
observational parameters as the source.  We therefore set the flux density 
scale using a value of 1.3 Jy for the phase calibrator, J1751-2524 (VLA online 
calibrator 
manual\footnote{\url{http://www.aoc.nrao.edu/\~{}gtaylor/calib.html}}).  The 
flux density of J1751-2524 at B and C configuration, determined by comparison 
with the primary flux density calibrator, varies from 1.3 Jy by as much as 
0.05 Jy depending on the frequency.  Therefore, a possible calibration error 
of up to 4\% was introduced into the 20 cm A-configuration data.

To determine the absolute polarization position angle, a comparison was made 
between polarization images made separately for A- and B-configuration data, 
tapered to the same resolution.  Forcing the angles to be the same in the high 
signal-to-noise regions resulted in a possible position angle error of up to 
3$\degr$ for the 20 cm, A-configuration data.

The second problem was at D configuration where there was insufficient 
parallactic angle coverage to solve for the instrumental and source 
polarization terms for the phase calibrator.  In this case, the instrumental 
polarization solutions were obtained by assuming that the flux density and 
phase calibrators had the same polarization as when they were observed in B 
and C configuration.  The rms phase errors for the polarization calibrator 
were less than 5$\degr$ after this correction.

After initial calibration, multiple passes of self-calibration were performed 
to improve the antenna phase solutions.  For the 6 cm data, the first few 
passes of self calibration at B configuration (the largest configuration size 
at this frequency) were done using a shorter spatial frequency range.  The 
final pass of self-calibration used the entire spatial frequency range and 
included an amplitude calibration.  In order to facilitate comparison between 
6 cm and 20 cm, the final 6 cm images were made with a similar beam shape and 
size as that at 20 cm and the maximum spatial frequency range was set to that 
at 20 cm.  The AIPS maximum-entropy deconvolution routine VTESS, which 
maximizes smoothness in an image within the confines of the data, was used to 
restore the total intensity images using a zero-spacing flux density estimate 
of 6.3 Jy which was the total flux density reported by \citet{dsa88} for 
their 6 cm image.  The default image for VTESS was a 6 cm image convolved to 
C-configuration resolution.  The standard correction for primary beam 
attenuation was applied to the final 6 cm images.  This correction raised the 
flux density at the outermost edge of the remnant by 6\%.

For the four frequencies at 20 cm, a shorter spatial frequency range was again 
used to self calibrate the A-configuration data before doing a pass using the 
entire spatial frequency range.  A final calibration was done in phase only 
using a flux-density-scaled 6 cm image.  Although self-calibration improves 
antenna phase, absolute position information is degraded.  Using the 6 cm 
image to calibrate in phase effectively registers one image to the other.  To 
solve the problem of an extended plateau in the beam pattern at low levels 
\citep{mld84}, the spatial frequency coverage weighting parameters in IMAGR 
were adjusted to provide a clean Gaussian beam.  The minimum spatial frequency 
range was also set to that at 6 cm.  The final 20 cm images were deconvolved 
using VTESS with the same default as that used at 6 cm but with the flux 
density scaled to 20 cm.  This essentially requires VTESS to make the 20 cm 
image look as much like the 6 cm image as possible with any resulting 
differences being due to the requirements of the data.  Variations in spectral 
index then result from significant differences in the data between 6 cm and 20 
cm instead of the reconstruction algorithm.  A zero-spacing flux density 
estimate of 17 Jy was used for image reconstruction in VTESS.  This value was 
also chosen to match that used by \citet{dsa88} who added the 17 Jy to their 
data before image construction.  The final images at 20 cm were corrected for 
primary beam attenuation which raised the flux density at the outer edge of 
the remnant by 0.5\%.

The AIPS routine UTESS, which maximizes emptiness in an image within the 
confines of the data, was used to restore Stokes Q and U images at 6 cm and 20 
cm.  The input images were made using IMAGR with the same spatial frequency 
coverage and weighting as used for the total intensity images.  No default 
images or zero-spacing flux density estimates were used for reconstruction.  
The final images were corrected for primary beam attenuation.

The images have been convolved to a 7$\farcs2$ round beam for several 
reasons.  First, Kepler's SNR is very far south for the VLA ($\delta_{2000} 
\approx -21\fdg5$) so the north-south antenna spacings are foreshortened to 
produce a very elliptical beam.  Second, in trying to make the 20 cm beam 
Gaussian, the beam size had to be increased resulting in a major axis of 
nearly 3$\farcs6$.  Third, even with a 3$\farcs6$ round beam, the 20 cm 
images had a rippling feature similar to sidelobe emission from a distant 
bright source which could not be eliminated.  Finally, using a larger beam 
increased the signal-to-noise ratio.  The higher resolution images were used 
in our proper motion study (B. Koralesky, L. Rudnick, \& T. DeLaney, in 
preparation).  

\section{Radio Images}
\label{sec:radim}

\subsection{Total Intensity}
\label{sec:totint}

\begin{figure}[ht]
\epsscale{1}
\plotone{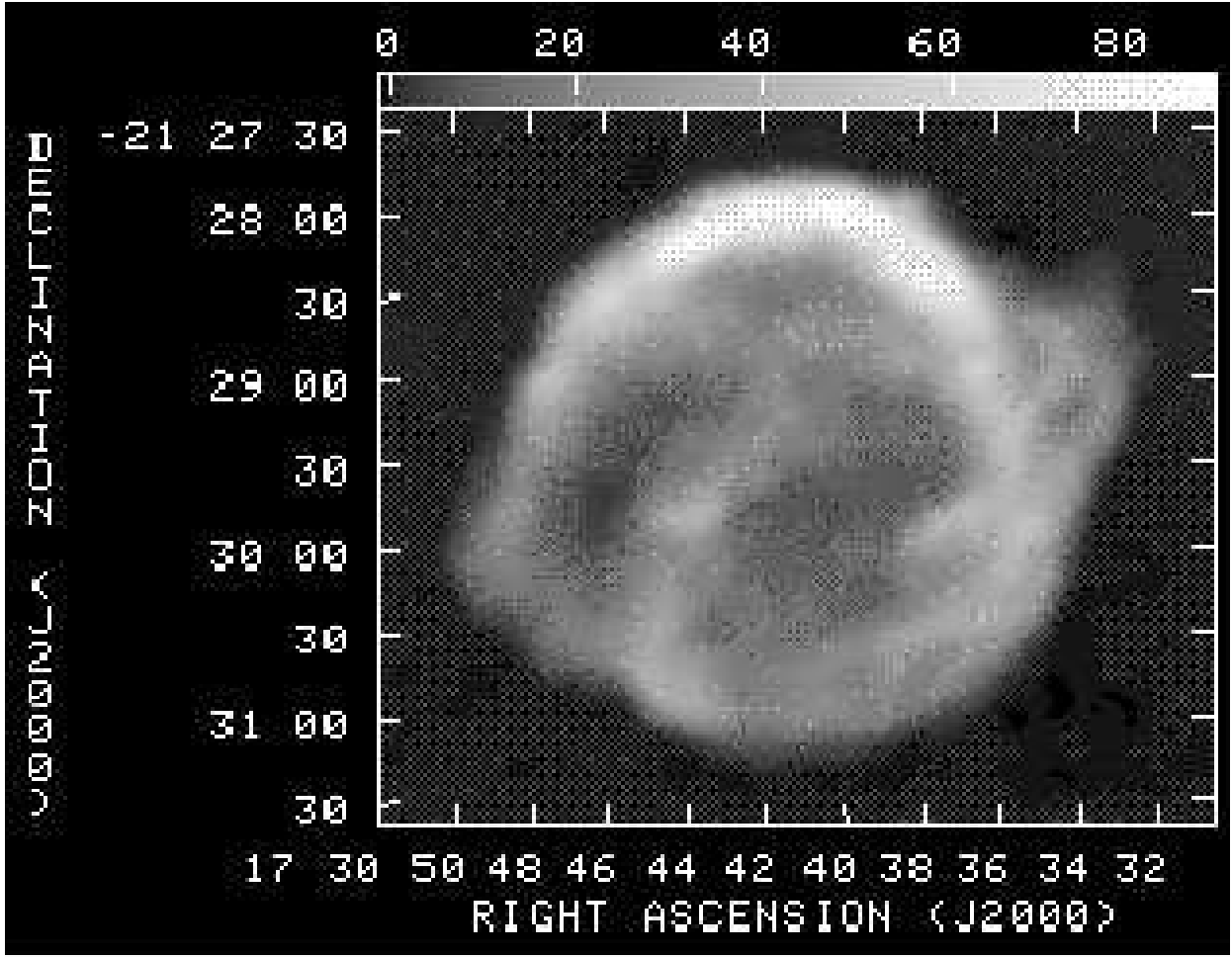}
\caption{Gray-scale image of Kepler's SNR at 20 cm.  The resolution is 
7$\farcs$2.  The brightness scale is in mJy bm$^{-1}$. \label{lband}}
\end{figure}

Representative 20 cm (1285 and 1365.1 MHz combined as described in 
\S \ref{sec:spvar}) and 6 cm (4835.1 MHz) total intensity images are shown in 
Figures \ref{lband} and \ref{cband}, respectively.  Table \ref{tbl2} 
summarizes the integrated flux densities and rms noises of the final total 
intensity images.  The flux densities agree to within 5\% of flux densities 
interpolated from the data compiled by \citet{re92}.

The total intensity images at 6 cm and 20 cm show the same features described 
by \citet{mld84} and \citet{dsa88}.  Kepler's SNR is a shell type with the 
northern portion of the ring about 3 times brighter than the southern portion 
of the ring.  The thickness of the northern portion of the ring is 
$\approx25\arcsec$ and the radius of curvature of this bright ring deviates 
from circular symmetry in the northwest.  For reference, the Galactic plane 
is $6\fdg8$ southeast of the remnant running from approximately northeast to 
southwest.

We now define nomenclature that will be used elsewhere in this paper to refer 
to spatial regions in Kepler's SNR.  These regions are shown schematically on 
Figure \ref{cband}.  The ``northern ring'' and ``southern ring'' refer to the 
limb-brightened shell.  The protrusion separating the northern ring from the 
southern ring in the east is called the ``eastern ear'' and the 
similar protrusion in the west is called the ``western ear.''  The ``central 
band'' is the emission ridge extending from the southeast to the central part 
of the remnant and ``steep arc'' refers to the ridge of emission crossing 
from east to west in the northern half of the remnant.  The small protrusion 
at $-25\degr$ is called the ``bump'' and the interior feature located just 
north of the southern ring at $-115\degr$ is called the ``western stub.''

\subsection{Spectral Index}
\label{sec:spindx}

\subsubsection{Variations Across the Remnant}
\label{sec:spvar}

\begin{figure}[ht]
\epsscale{1}
\plotone{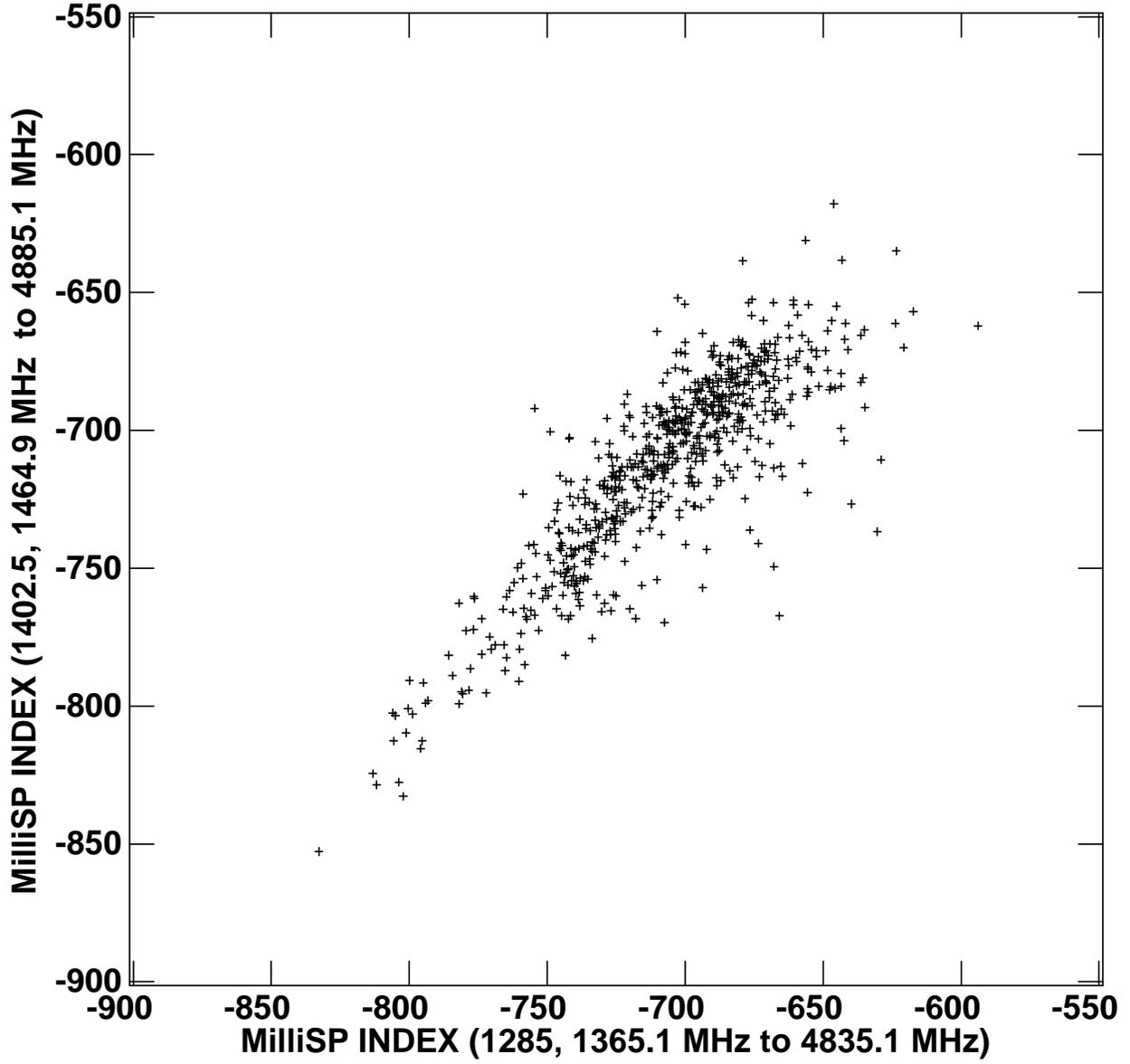}
\caption{Spectral index from pair 2 \emph{vs.} spectral index from pair 1.  
Points are sampled once per resolution element (beam). 
\label{a1va2}}
\end{figure}

\begin{figure}[ht]
\epsscale{1}
\plotone{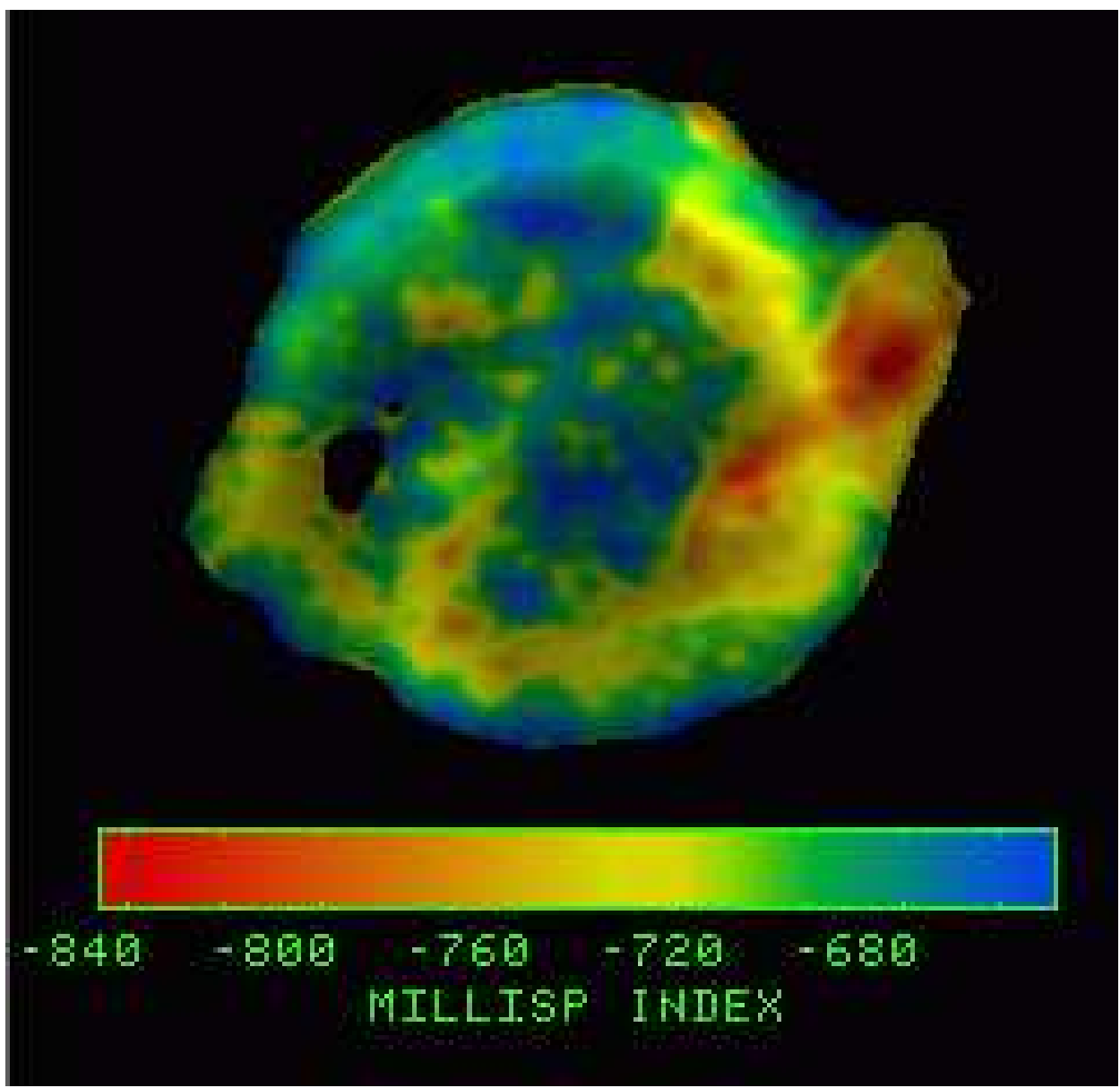}
\caption{Spectral index between 6 cm and 20 cm.  Intensity 
is set by the 6 cm continuum image. \label{spix}}
\end{figure}

For the spectral index measurements, two independent maps were made at 
each of the 20 cm and 6 cm bands.  The first spectral index map (pair 1) 
consisted of data from 1285 and 1365.1 MHz (20 cm), and 4835.1 MHz (6cm).
The second spectral index map (pair 2) consisted of data from 1402.5 and 
1464.9 MHz (20 cm), and 4885.1 MHz (6cm).  The differences in spectral index 
between the two independent images were small, except at low 
signal-to-noise.  These images were therefore blanked when the 20 cm 
flux density was less than 5 mJy bm$^{-1}$.  Figure \ref{a1va2} shows the 
consistency between the two spectral index measurements.  To eliminate 
oversampling, only values of pixels that are separated by one full beamwidth 
are displayed.  After blanking, the rms difference between the two 
independent spectral index images is $\pm$0.019.  The two spectral index 
images were then averaged together to make the final spectral index image 
which is shown in Figure \ref{spix}.  The color is determined by the 
spectral index and the intensity is set by the 6 cm (4835.1 MHz) continuum 
radio image.  

\begin{figure}[ht]
\epsscale{1}
\plotone{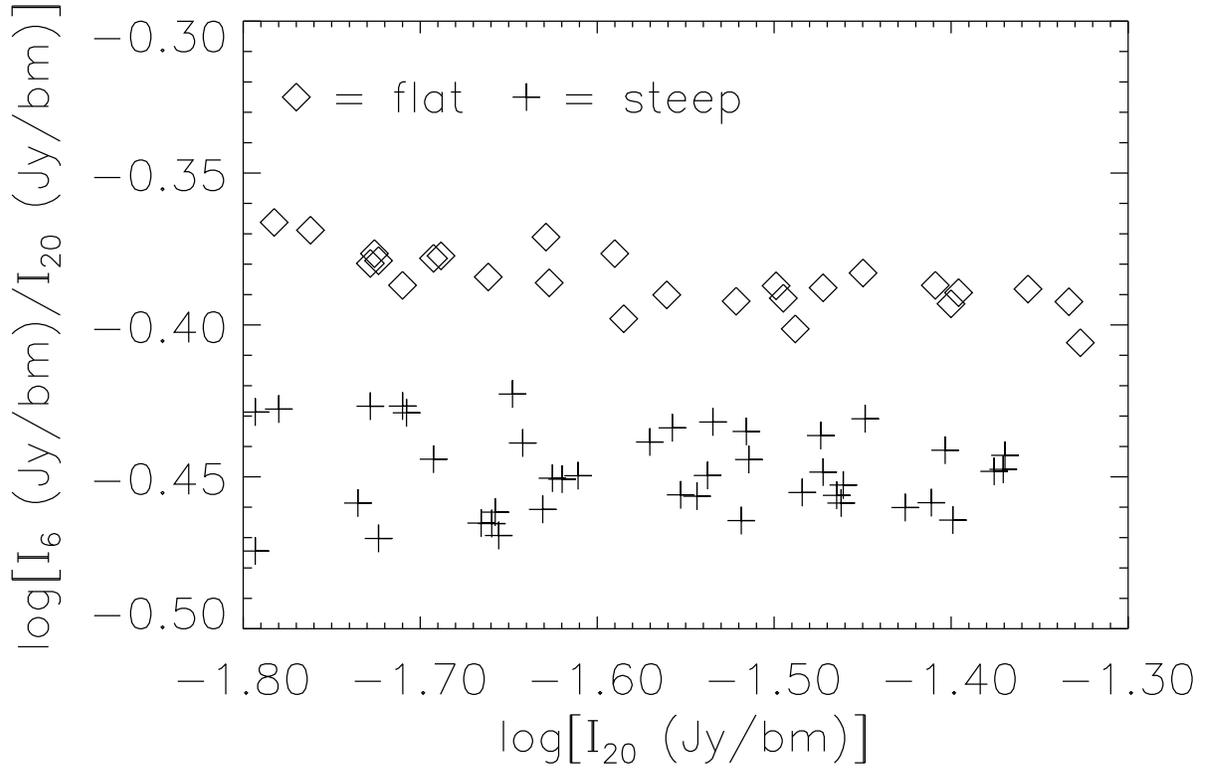}
\caption{log(I$_{6}$/I$_{20}$) \emph{vs.} Log(I$_{20}$), where 
I$_{\mathrm{x}}$ is the intensity in band x, for a continuous flat spectral 
region (diamonds) and a continuous steep spectral region (pluses). 
\label{spixlog}}
\end{figure}

For the first time reliable spectral index variations have been measured 
in the remnant between 6 cm and 20 cm.  The spectral index ranges from $-$0.85 
to $-$0.6. However, the spectral index in Figure \ref{spix} ranges from 
$-$0.85 to $-$0.65 to better enhance the details.  To assess the reliability 
of the variations in the final spectral index image, a continuous flat 
spectral region in the northern ring and a continuous steep spectral region 
in the western ear and western stub were chosen from the total intensity 
images (pair 1).  The regions chosen are about 2700 arcsec$^2$ in area 
representing about 50 independent beams each.  Figure \ref{spixlog} is the 
plot of log(I$_{6}$/I$_{20}$) \emph{vs.} log(I$_{20}$) where I$_{\mathrm{x}}$ 
is the intensity in band x for each region.  There is a distinct separation 
between the flat and steep components, much greater than the scatter and 
showing no significant biases as a function of intensity.

\begin{figure}[ht]
\epsscale{.60}
\plotone{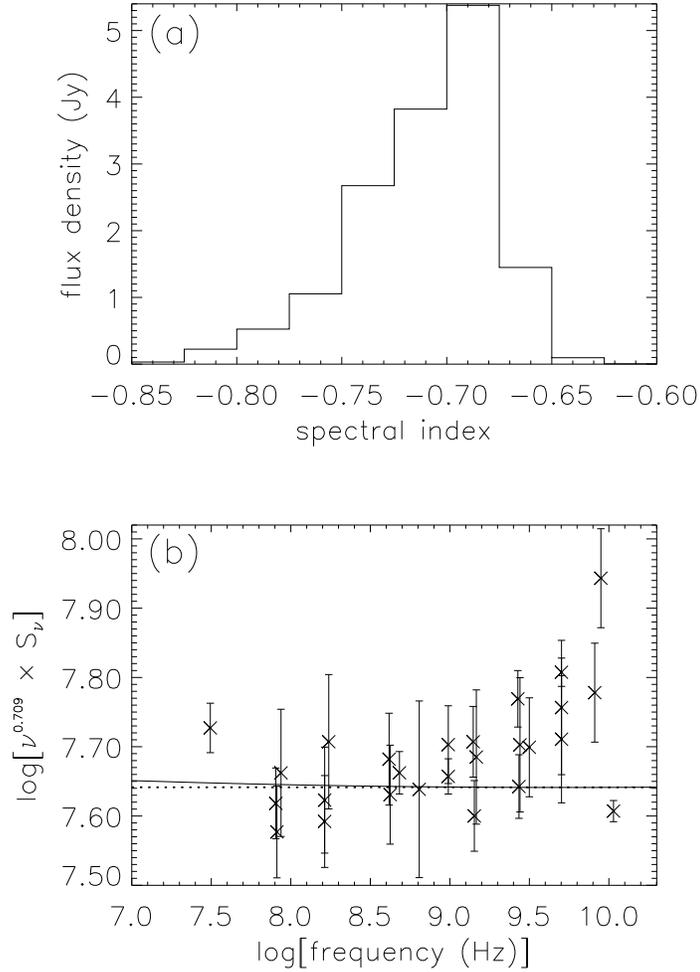}
\caption{(a) The result of binning the observed 20 cm flux densities 
into 10 spectral index bins.  (b) The solid line is the result of adding 
together the 10 power laws representing each bin.  The dotted line represents 
a power law with an index of $-$0.709.  The data points with errors are taken 
from a spectral plot by \citet{re92}.  \label{reynolds}}
\end{figure}

As a test to determine if the measured spectral index variations could 
reproduce the observed small curvature in the integrated spectrum 
\citep{re92}, the total 20 cm (using pair 1) flux density at each spectral 
index was measured and binned, as shown in Figure \ref{reynolds}a.  These 
binned data were then used to predict the integrated flux densities at other 
frequencies by summing the flux density contributions determined for each 
spectral index box assuming each spectral index bin represented a different 
power law fixed to its observed flux density at 20 cm.  The predicted 
spectrum is shown as the solid line in Figure \ref{reynolds}b.  The data 
points and error bars in Figure \ref{reynolds}b were taken from \citet{re92} 
and trace out the observed integrated spectrum.  Although the spectrum 
predicted from our superposed power laws model shows small positive 
curvature, it is insufficient to describe the current data between 10 MHz and 
10 GHz.  Thus there must be some mechanism, such as the cosmic ray 
modification of shocks proposed by \citet{re92}, that produces the positive 
curvature in the spectrum.

\subsubsection{Mean Spectral Index}

Although we are primarily interested in spectral index variations, for 
completeness we have also determined the mean spectral index.  We use two 
methods.  The first is to determine the mean value in the spectral index 
image which yields a value of $-$0.71 with a formal error in the mean of 
0.001.  The second method uses the integrated flux densities at 6 cm and 
20 cm (using the mean spectral index value from pair 1 and pair 2)
and yields the same value.  The actual errors in the mean spectral index are 
dominated by calibration errors, yielding an estimated uncertainty of 
$\approx$ 2\% and by the missing flux density from the interferometer's lack 
of sensitivity to the largest scale emission.  Since the spatial frequency 
sampling of the 6 cm and 20 cm images were matched for our spectral index 
analysis, we expect the same degree of missing flux density at each 
waveband.  We cannot assign an error to the missing flux density because 
modern single dish flux density measurements are not available.  Our 
results differ from the mean spectral index of $-$0.65 
reported in \citet{dsa88}.  We believe the earlier results are 
less reliable because of their poorer available interferometer sampling, and 
the use of data from only the bright northern ring, which we found to be 
flatter than average.

\subsubsection{Spectral Tomography}
\label{sec:sptomo}

Because the radio emission is optically thin, it is possible that some of 
the measured spectral variations can be confused by overlapping structures.  
Spectral tomography is a technique designed to separate such structures.  
This technique has been applied to analyze the spectra of radio galaxies 
\citep{kr97} and, recently, to Tycho's SNR \citep{kkl00}.  Because the 
spectral index of the overlapping components is not known, a gallery of 
images is made each with a specific spectral component subtracted out.  
Mathematically, a set of images $M_t(\alpha_t)$ is constructed where,
\begin{displaymath}
M_t(\alpha_t) \equiv M_{20} - M_{6}\left(\frac{\nu_{20}}{\nu_{6}}\right)^
{\alpha_t},
\end{displaymath}
where $M_{6,20}$ are the images at 6 cm (4835.1 MHz) and 20 cm (using pair 
1).  If a component has spectral index $\alpha_t$, it will disappear with 
respect to its surroundings in image $M_t$.  If a component has spectral 
index $>\alpha_t$, it will be oversubtracted and have negative brightness.  
If a component has spectral index $<\alpha_t$, it will be undersubtracted and 
have positive brightness.  Note that the tomography images are intensity 
weighted as well as showing spectral index changes.  

\begin{figure}[ht]
\epsscale{1}
\plotone{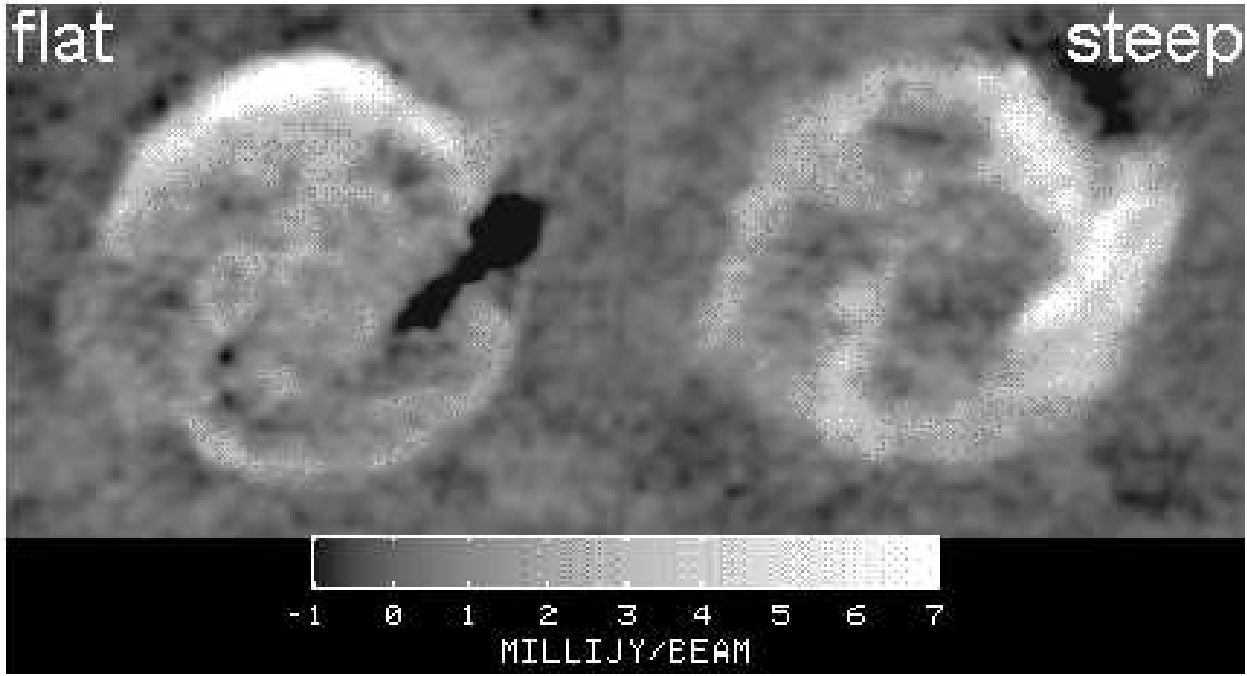}
\caption{Two representative spectral tomography images between 20 cm and 
6 cm.  On the right, bright areas have radio spectral indices steeper 
than -0.65.  On the left, the intensity scale is reversed and the zero point 
shifted so that bright areas have spectral indices flatter than -0.75 and the 
dark areas steeper indices.  The grey-scale for intensity is the same on both
images. \label{tomo}}
\end{figure}

Figure \ref{tomo} consists of two images from the tomography gallery.  The 
image on the left (hereafter called flat) is the result of setting 
$\alpha_t=-$0.75 and represents a flat component of emission.  For display 
purposes, this image has been reversed so that what is actually shown is 
$-M_{-0.75}$; therefore bright spatial components are flatter in spectral 
index than this value.  Essentially all of the steep emission has been 
subtracted out to give zero or negative brightness values at the locations of 
those features, with the steepest features in the west substantially 
oversubtracted.  What shows up as positive brightness results from emission 
flatter than $\alpha_t=-$0.75.  The image on the right (hereafter called 
steep), for $\alpha_t=-$0.65, has the flatter component subtracted out so 
that positive brightness represents emission steeper than $\alpha_t=-$0.65 
and negative brightness represents flatter emission.  

Prominent features on the flat-spectrum radio image include the northern ring 
and a thin southern ring.  Many of the interior features have disappeared and 
the eastern ear is very faint.  The western side of the remnant (western ear 
and western stub), the bump, and parts of the central band have been 
oversubtracted in the flat image but are prominent in the steep-spectrum 
radio image.  The eastern ear is relatively more defined in the steep image 
than in the flat image.  Parts of the northern and southern ring are also 
steep and the steep arc is clearly defined in the steep-spectrum radio 
image.  When we refer to structures as flat, they appear prominently in the 
flat image and they are predominantly green or blue in Figure \ref{spix}.  
Similarly, when we refer to structures as steep, they appear prominently in 
the steep image and they are predominantly yellow, orange, or red in Figure 
\ref{spix}.  It would be an oversimplification to state that the remnant 
consists of exactly two overlapping spectral components.  Many areas of the 
remnant show a spectral index gradient indicating that the remnant is likely 
composed of multiple overlapping spectral index structures rather than just 
two structures.  

\subsection{Polarization Results}
\label{sec:pol}

\subsubsection{Polarized Intensity}

\begin{figure}[ht]
\epsscale{1}
\plotone{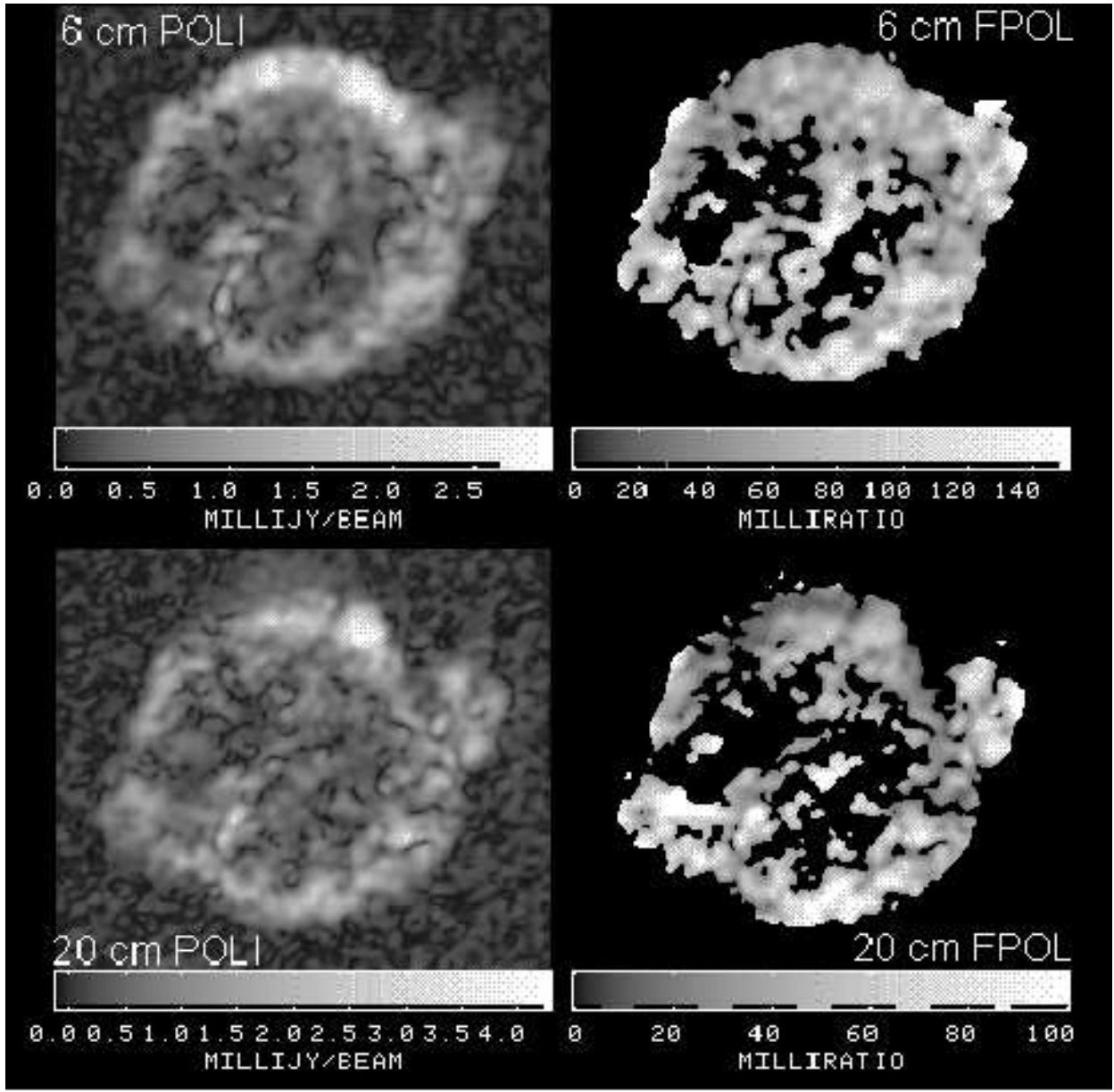}
\caption{Gray-scale images of 6 cm polarized intensity (top left), 
6 cm fractional polarization (top right), 20 cm polarized 
intensity (bottom left), and 20 cm fractional polarization (bottom 
right). \label{poln}}
\end{figure}

Noise corrected linearly polarized intensity (POLI) images for one frequency 
at 6 cm (4835.1 MHz) and one frequency at 20 cm (1365.1 MHz) were made from 
the Stokes Q and U images using the AIPS routine COMB.  These images, and the 
fractional polarization (FPOL) images are shown in Figure \ref{poln}.  The 
fractional polarization image at 6 cm was blanked based on intensity values 
below 0.5 mJy bm$^{-1}$ on the 6 cm total intensity image while the 20 cm 
fractional polarization image was blanked below 2 mJy bm$^{-1}$ on the 20 cm 
total intensity image.  The northern and southern rings and ears are clearly 
identified in the polarization intensity images although the north-south 
brightness asymmetry is not as pronounced as it is in the total intensity 
images.  The interior structures, excepting the central band, are not at all 
defined and the bump is also difficult to discern in the polarized intensity 
images.

In order to determine the mean percent polarization in the remnant, two 
definitions will be used because they weight the polarized emission 
differently.  The first is $\langle$P$_{\mathrm{x}}\rangle$ where the 
P$_{\mathrm{x}}$ are the pixel values from the fractional polarization images 
for band x in Figure \ref{poln}.  Using this definition, 
$\langle$P$_{6}\rangle=6.1\pm$0.1\% while 
$\langle$P$_{20}\rangle=3.8\pm$0.1\%.  The second method for determining the 
``mean'' percent polarization is to calculate 
$\Sigma$POLI$_{\mathrm{x}}/\Sigma$I$_{\mathrm{x}}$ where POLI is polarized 
intensity, the x's are defined as before, and the sums refer to integrated 
intensities on unblanked images.  Based on this definition, 
$\Sigma$POLI$_{6}/\Sigma$I$_{6}=6$\% and 
$\Sigma$POLI$_{20}/\Sigma$I$_{20}=3.5$\% which are close to the values 
measured by $\langle$P$_{6}\rangle$ and $\langle$P$_{20}\rangle$.  In 
comparison, \citet{mld84} report 
$\Sigma$POLI$_{\mathrm{x}}/\Sigma$I$_{\mathrm{x}}=4.4$\% and 2.2\% for 6 cm 
and 20 cm, respectively.  We have examined a number of possibilities for why 
our values differ from those of \citet{mld84} but we cannot identify the 
cause of the discrepancy without reevaluating the 1984 data.

The degree of polarization is an important measure in SNRs because it is an 
indicator of the magnetic field ordering and the ratio of fractional 
polarizations at different bands is a measure of the mixing of synchrotron and 
thermal plasmas.  For a uniform magnetic field, the degree of polarization of 
a synchrotron source is independent of frequency and is directly related to 
the spectral index by P$=(3-3\alpha)/(5-3\alpha)$.  For Kepler's mean 
spectral index of $-$0.71, the mean percent polarization should be about 
72\%.  In order to account for the low percent polarization in Kepler's SNR, 
as well as other young SNRs \citep{mil87}, the magnetic field must have a 
high degree of disorder on subarcsecond scales.

\subsubsection{Depolarization}

\begin{figure}[ht]
\epsscale{1}
\plotone{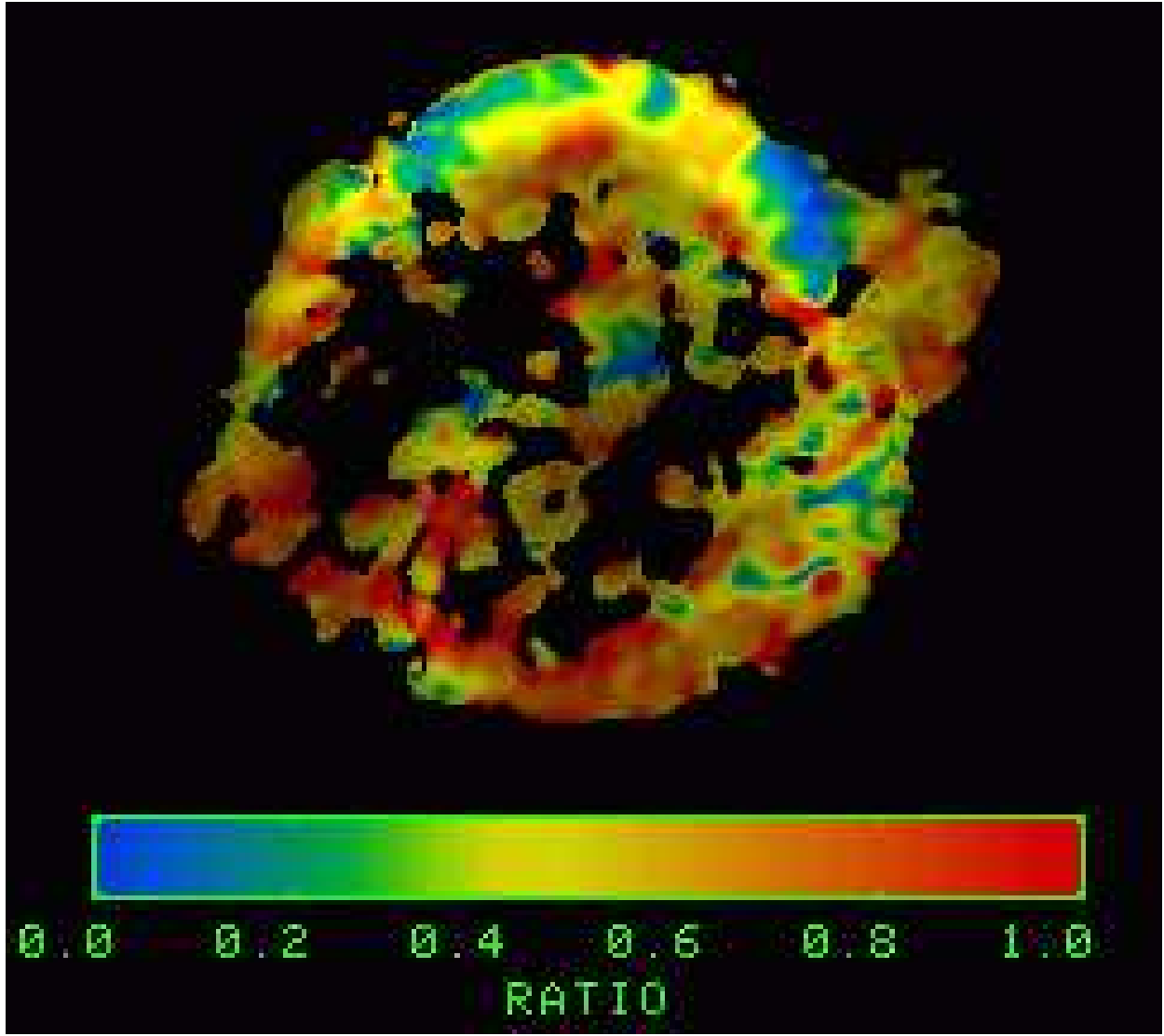}
\caption{Depolarization image from 6 cm to 20 cm as defined 
in the text.  Intensity is set by the 6 cm radio total intensity 
image. \label{depol}}
\end{figure}

Another effect that further depolarizes SNRs at longer wavelengths is 
differential Faraday rotation of emission from regions at different depths in 
the SNR.  Figure \ref{depol} is a depolarization (P$_{20}/$P$_{6}$) image 
showing variations over the full range of 0 (complete depolarization) to 1 (no 
depolarization).  The image has been blanked based on the 6 cm polarized 
intensity of 0.3 mJy bm$^{-1}$ below which noise becomes important.  A 
north-south asymmetry is apparent here as portions of the bright northern ring 
are strongly depolarized while the southern ring shows little depolarization.

\subsubsection{Faraday Rotation and Magnetic Field}
\label{sec:rmb}

Because the Faraday rotation is dependent on $\lambda^2$, we can determine 
it by combining the position angles of the polarization vectors at multiple 
wavelengths.  To do this, average Stokes Q and U images were calculated at 20 
cm (using the 1365.1, 1402.5, and 1464.9 MHz images) and at 6 cm (using both 
frequency images), and then used to construct 20 cm and 6 cm polarization 
angle images.  An angle difference image was constructed between 20 cm and 6 
cm.  The angle difference image was blanked based on the 20 cm polarized 
intensity of 0.4 mJy bm$^{-1}$ below which noise becomes important and was 
then used to calculate the rotation measure image.  The angle difference 
image had to be corrected for 180$\degr$ ambiguities.  To do this, a critical 
assumption was made that there should be no abrupt, large angle difference 
jumps from one pixel to the next in the image.  Therefore, small scale 
``jump'' regions were shifted 180$\degr$ to conform to the larger surrounding 
regions.  By constraining the image to have smooth transitions between 
regions, the total range of angle differences is found to be larger than 
180$\degr$.  The procedure of assuming no sudden jumps of 180$\degr$ appears 
to produce a unique result, except for an overall 180$\degr$ ambiguity for 
the entire image.  This ambiguity is removed by looking at the angle 
difference between two 20 cm (1365.1 and 1464.9 MHz) polarization angle 
images.  A rotation measure image was also made using the AIPS routine RM, 
which does not involve application of the no-jump criterion.  The resulting 
image had only a few minor differences from the manually constructed rotation 
measure image after appropriate blanking for noise.

\begin{figure}[ht]
\epsscale{1}
\plotone{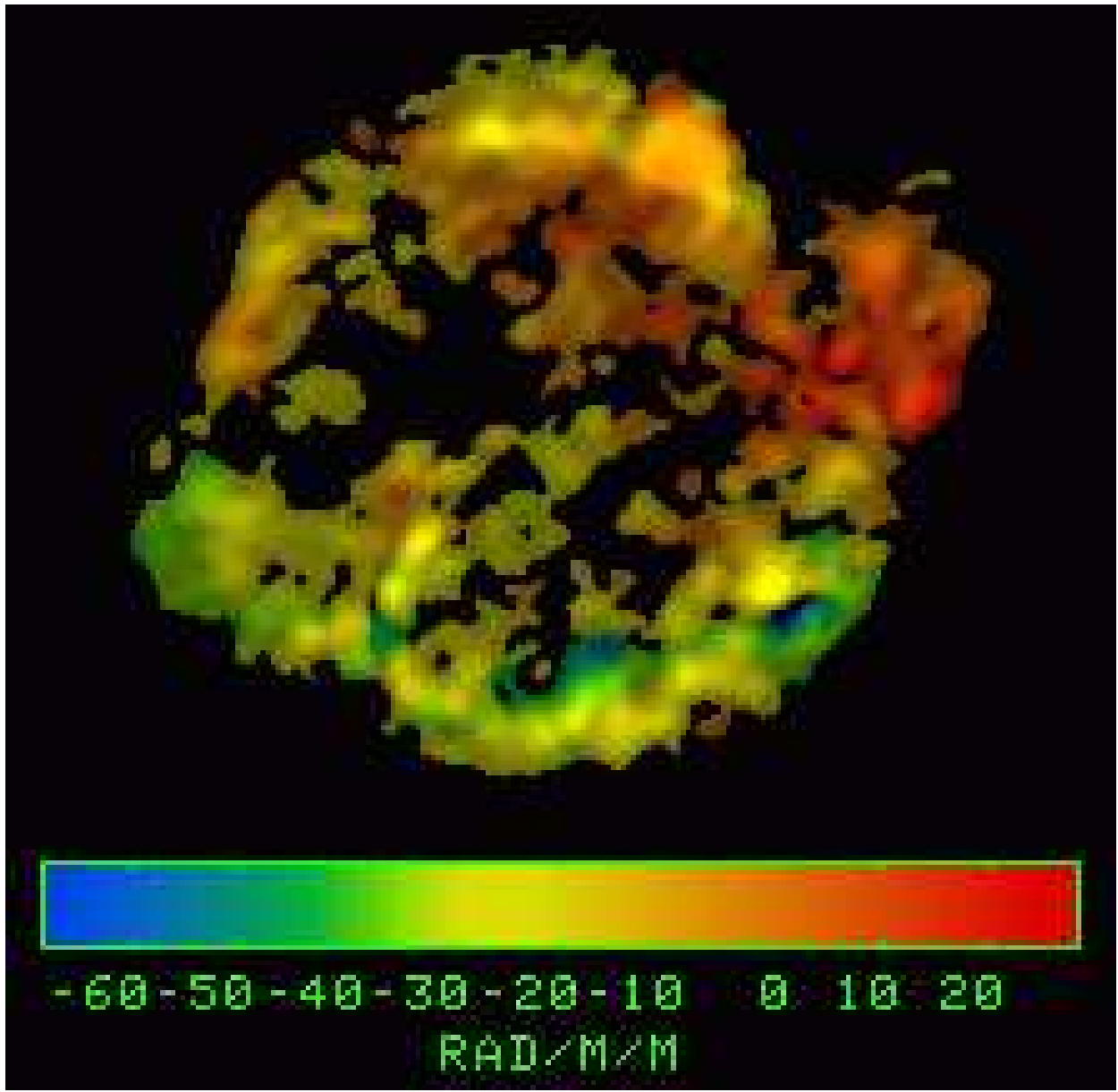}
\caption{Rotation measure between 6 cm and 20 cm.  Intensity is 
set by the 20 cm polarized intensity image. \label{rm}}
\end{figure}

The final rotation measure image is shown in Figure \ref{rm} and reveals the 
same north-south asymmetry seen elsewhere.  The western ear has more positive 
rotation measures than the rest of the remnant while the southern ring has the 
most negative rotation measures in the remnant.  There is also a steep 
gradient in rotation measure across the bump.  These results correspond well 
with those presented in \citet{mld84}, although they do not allow variations 
larger than 180$\degr$ across their image.  Since we do not expect the 
external Faraday screen to vary substantially over the angular size of 
Kepler's SNR ($\approx 200\arcsec$) \citep{sk80}, the range of rotation 
measure from $-$67.7 rad m$^{-2}$ to 27.7 rad m$^{-2}$ is probably the result 
of variations in plasma density and thickness and magnetic field strength and 
orientation within the remnant as pointed out by \citet{mld84}.

\begin{figure}[ht]
\epsscale{1}
\plotone{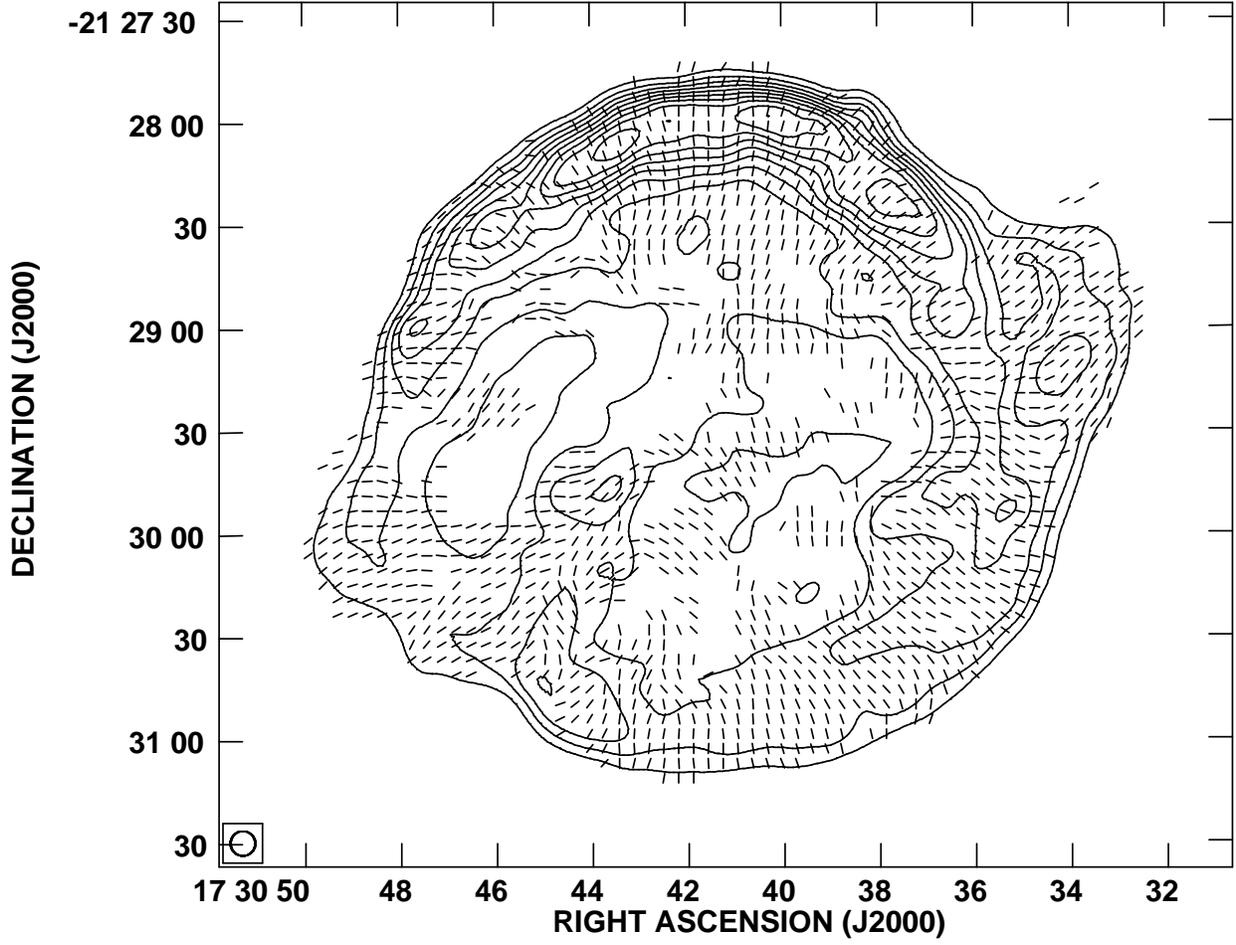}
\caption{The magnetic field structure in Kep\-ler's SNR.  All vectors are the 
same length.  The contours are 6 cm total intensity. \label{bfield}}
\end{figure}

Figure \ref{bfield} shows the magnetic field structure in the remnant, after 
correction for the derived rotation measure.  All vectors have been set to the 
same length.  The contours are 6 cm radio continuum.  The magnetic field is 
radial for the most part, as seen in other young SNRs \citep{mil87},  but 
there are significant deviations from radial on scale sizes of $\approx 
20\arcsec$.  In general, there is more disorder in the south than the north 
as noted by \citet{mld84} for their polarization angle images.  Recent proper 
motion measurements show many of the same patterns as the magnetic field 
structure (B. Koralesky, L. Rudnick, \& T. DeLaney, in preparation).

\section{Optical, Infrared, and X-ray Images}
\label{sec:oirx}

An H$\alpha$ + [\ion{N}{2}] (hereafter just H$\alpha$) image was obtained by 
\citet{blv91} and kindly provided to us by W. Blair.  This image was taken in 
1987 with the 2.5 m DuPont telescope at Las Campanas.  The stars on this image 
were already removed to first order by subtracting a continuum image.  There 
were enough intact stellar remnants, though, to register the H$\alpha$ image 
to the stars on a DSS image to an accuracy of better than 1$\farcs5$.  

\begin{figure}[ht]
\epsscale{1}
\plotone{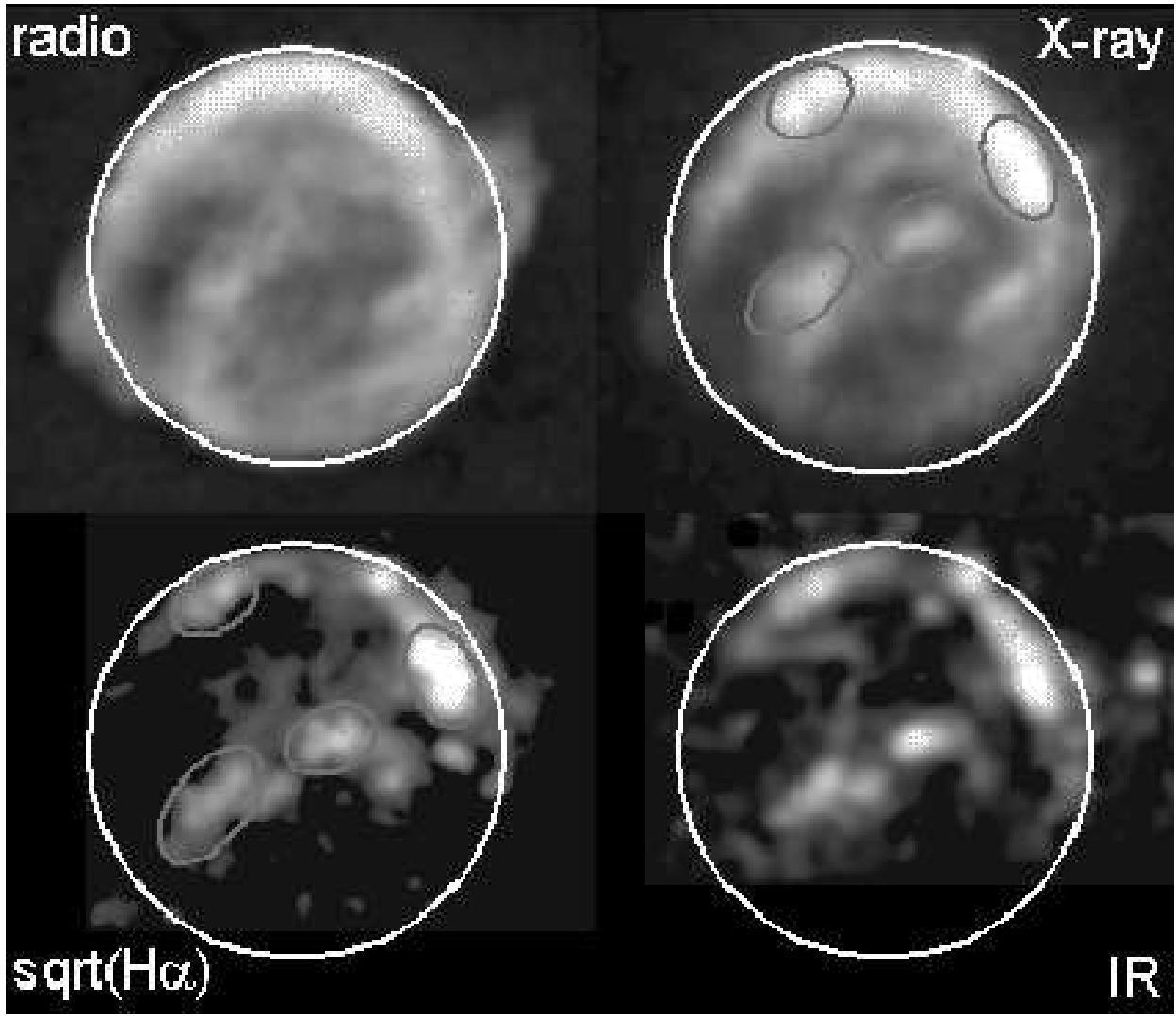}
\caption{Gray-scale images of 6 cm radio continuum (top left), 
X-ray (top right), H$\alpha$ (bottom left), and IR (bottom right).  The 
circles are 100$\arcsec$ in radius.  \label{cxoi}}
\end{figure}

In order to compare the distribution of H$\alpha$ to our radio images at a 
resolution of $7\farcs2$, we had to take an extra step to further remove 
significant contributions from stars.  The stars embedded in the H$\alpha$ 
emitting regions were identified by detailed comparisons with images by 
\citet{dbd86} and \citet{bb91} and ``removed'' by replacing the pixel 
intensities with that of nearby, diffuse H$\alpha$ emission in the same 
regions.  To remove stars that were not embedded in emission, the local mean 
was determined for a region immediately adjacent to each star.  These stars 
were then ``removed'' by replacing their pixel intensities with the local 
mean.  Our final convolved H$\alpha$ image is shown in Figure \ref{cxoi} with 
a square root transfer function to enhance the faint emission.

Although the H$\alpha$ image is ten years older than those at the other 
wavelengths, this is not a problem because of the small proper motions of 
H$\alpha$ features.  The fastest moving knot in right ascension would have 
moved 0$\farcs384$ and the fastest moving knot in declination would have 
moved 0$\farcs367$ in ten years \citep{bb91}.  These are well within the 
registration errors.

An IR image was obtained by \citet{dlb99, dlc01} and kindly provided to us by 
P.-O. Lagage and T. Douvion.  This image represents data taken in 1996 with 
the LW8 filter (10.7-12 $\micron$) aboard ISO.  The IR image was registered 
in the same manner as the H$\alpha$ but using a 2MASS image as the template 
and yielding, again, a positional uncertainty of 1$\farcs5$.  Due to the 
resolution of ISO (6$\arcsec$), individual stars in the emitting regions 
could not be isolated and removed.  The final image is shown in Figure 
\ref{cxoi} and was convolved to 7$\farcs2$ for comparison to the other 
wavelength images.

To make the X-ray image shown in Figure \ref{cxoi}, the ROSAT high resolution 
imager data taken in 1997 and used by \citet{hug99} were obtained from the 
ROSAT public archive.  An image was made at 1$\arcsec$ pixel$^{-1}$ using the 
program FTOOLS and then convolved with a 7$\farcs2$ beam.  The registration 
of the X-ray image is not a straightforward process.  Unlike the IR, there 
are no groundbased observations at X-ray energies to establish an absolute 
coordinate system.  It is also unclear exactly what the spatial relationship 
is between the X-rays and emission at other wavelengths.  Typically, X-ray 
images have been registered to radio images because of the similar 
distribution of the diffuse emission and the general similarity of the 
locations of large-scale bright regions.  A comparison of the X-ray and 
H$\alpha$ images, though, clearly shows several features common to both such 
as the emission at the center, in the northwest, and the northeast (outlined 
in red on Figure \ref{cxoi}).  Given these specific regions of similarity, 
the X-ray image was registered by maximizing the correlation of emission with 
the H$\alpha$ image.  This was done by spatially shifting one image with 
respect to the other and assuming that there were no rotation or scaling 
differences between the two images.  A difference image was constructed 
for each position shift.  The final shift position was set when the residuals 
for each of the ``similar'' regions were minimized.  The positional 
uncertainty between the two images is 1$\farcs$5.  To determine how different 
the position of the X-ray image could have been if we had registered to a 
radio image instead, angle averaged radial brightness profiles for the radio 
total intensity and X-ray images were created using the same parameters as the 
comparison by \citet{mld84} who registered their Einstein HRI image to their 
radio total intensity image.  The relative positions of our X-ray and radio 
total intensity images agree to within 2$\arcsec$ with the Einstein HRI and 
radio registration of \citet{mld84}.

The X-ray and H$\alpha$ images show a general north-south brightness 
asymmetry.  Although the IR image has no data in the south, IRAS 
observations \citep{bra87} indicate the same north-south brightness asymmetry 
as seen in the other wavebands.  The H$\alpha$ and IR images are remarkably 
similar.  Both images are brightest in the northwest, have a ``gap'' in the 
north corresponding to reduced surface brightness, have emission 
corresponding to the central band, have a clearly defined bump, and have 
emission slightly west of center.  The reader is reminded that the H$\alpha$ 
image includes [\ion{N}{2}] emission and, in fact, spectra taken by 
\citet{blv91} indicate that the bright northwest region is probably dominated 
by [\ion{N}{2}] emission.  The X-ray image has emission which corresponds to 
the emission on the H$\alpha$ and IR images.  However, there are regions of 
dissimilarity as well.  Unlike the H$\alpha$ and IR, the X-ray emission has a 
much less pronounced gap in the north and the ears and southern ring are 
clearly defined (with the caveat that there are no data in the south for the 
IR image).  The reader is also reminded that the comparison here involves 
X-rays at energies $\lesssim$ 2 keV.  No statement is implied about the 
distribution of X-ray emission at higher energies.

\section{Comparisons}
\label{sec:comps}

The most striking aspect of all of the images presented in this paper is the 
north-south asymmetry.  This is seen in brightness (Figure \ref{cxoi}), 
spectral index (Figure \ref{spix}), polarization (Figure \ref{poln}), 
depolarization (Figure \ref{depol}), and rotation measure (Figure \ref{rm}).  
Other interesting features include the bump which is distinctly visible in 
the brightness images in all wavebands, has a steep radio spectrum, and has a 
rotation measure gradient.  The very steep western ear is not prominent in 
either the H$\alpha$ or IR images, is not depolarized, but has the most 
positive rotation measures.  In contrast, the somewhat steep eastern ear has 
negative rotation measures but shares all the other properties of the western 
ear.

\begin{figure}[ht]
\epsscale{1}
\plotone{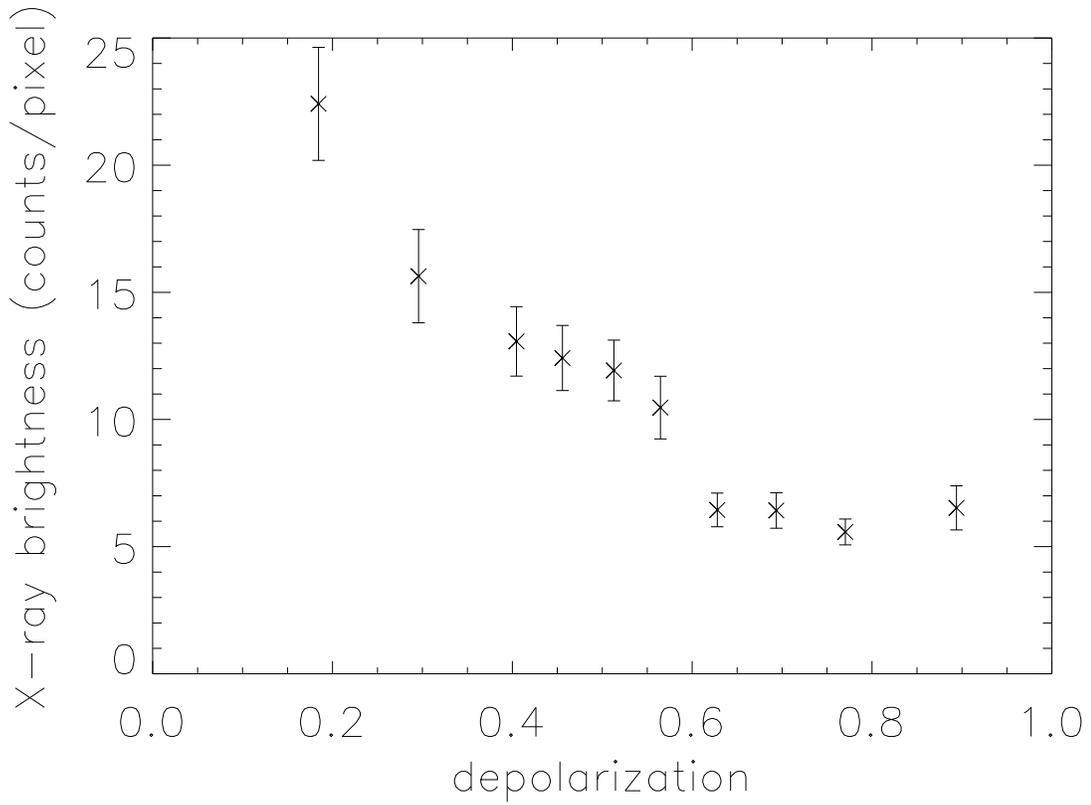}
\caption{Comparison of  X-ray brightness \emph{vs.} depolarization.  The 
$\times$'s represent the mean X-ray brightness in bins which are 45 
beam-independent data points wide in depolarization.  The error bars 
represent the rms error in the mean X-ray brightness in each depolarization 
bin.  The rms scatter per independent beam is $\approx$ 7 times higher in each 
bin than the errors in the mean.
\label{depvx}}
\end{figure}

\begin{figure}[ht]
\epsscale{1}
\plotone{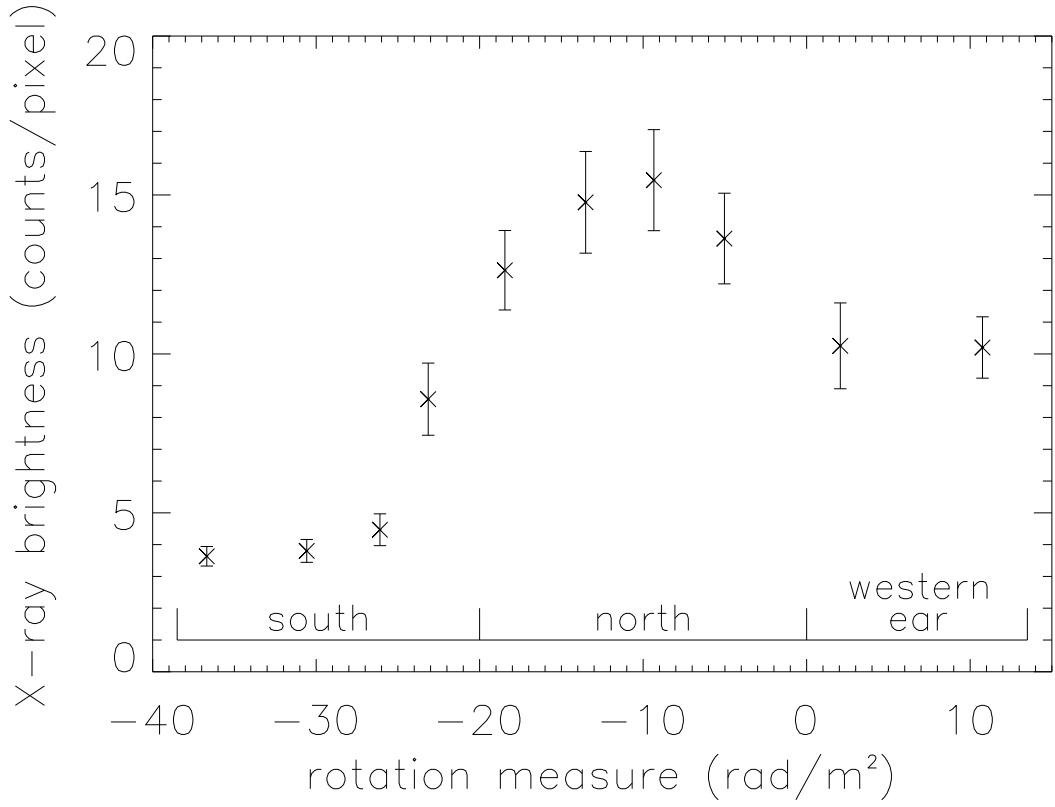}
\caption{Comparison of  X-ray brightness \emph{vs.} rotation measure.  The 
$\times$'s represent the mean X-ray brightness in bins which are 50 
beam-independent data points wide in rotation measure.  The error bars are set 
by the rms of the mean X-ray brightness in each bin.  
\label{rmvx}}
\end{figure}

The correlation between thermal emission and both depolarization and rotation 
measure is expected since Faraday rotation is proportional to the integral of 
thermal electron density times magnetic field along the line of sight and both 
bremsstrahlung and line emission are proportional to the square of the 
electron density.  The relation between depolarization and X-ray emission can 
be seen in Figure \ref{depvx}.  The data were binned by their depolarization 
with 45 independent beams per bin.  The $\times$'s and the error bars show 
the mean and error in the mean of the X-ray brightness in each depolarization 
bin.  Figure \ref{rmvx} is a similar plot for rotation measure and X-ray 
emission with the data binned by rotation measure into 50 independent beams 
per bin.  Data from the western ear dominates the region with rotation 
measures greater than 0 rad m$^{-2}$ while data from the southern ring 
dominates the region with rotation measures less than $-20$ rad m$^{-2}$.  
The region between $-20$ and 0 rad m$^{-2}$ contains data primarily from the 
bright northern ring and the less bright interior of the remnant.  If the 
western ear and the interior of the remnant are excluded, then there is a 
clear correlation between rotation measure and X-ray brightness as reported 
by \citet{mld84}.  To determine where zero rotation actually lies, so that it 
can be determined if more rotation corresponds to brighter X-ray regions, the 
galactic contribution to the rotation measure must be removed.  Because 
Kepler's SNR is about 600 pc above the galactic plane, we assume that all of 
the galactic Faraday screen in that section of the sky lies between us and 
Kepler.  Using the catalog of \citet{bmv88}, the median rotation measure of 
all extragalactic sources within 20$\degr$ of Kepler's SNR (21 sources) is 5 
rad m$^{-2}$ excluding two sources with rotation measure beyond 2$\sigma$.  
However, the rms of the remaining 19 sources is 77 rad m$^{-2}$ which is too 
large to allow a reliable determination of the galactic contribution to the 
rotation measures in Kepler's SNR.  There is thus no way to test whether 
X-ray brightness is related to more or less rotation.

\begin{figure}[ht]
\epsscale{.60}
\plotone{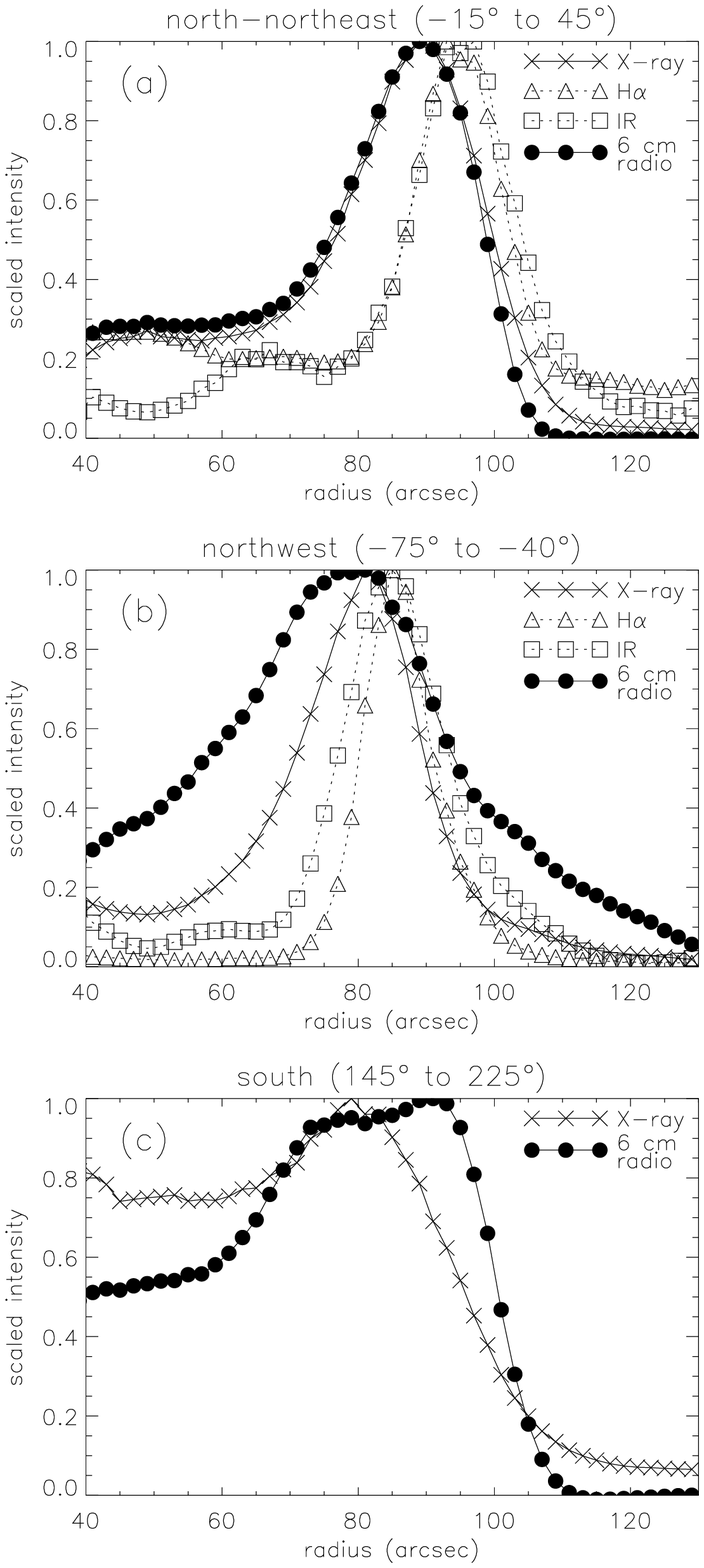}
\caption{Angle averaged radial profiles at all wavebands for (a) the north to 
northeast ($-15\degr$ to 45$\degr$), (b) the northwest ($-75\degr$ to 
$-40\degr$), and (c) the southern ($145\degr$ to $225\degr$) regions.  The 
center used for the azimuthal averaging  is $\alpha_{2000} = 17^{\mathrm{h}} 
30^{\mathrm{m}} 41\fs25$ and $\delta_{2000} = -21\degr 29\arcmin 29\farcs7$. 
\label{irings}}
\end{figure}

In Figure \ref{irings} we plot angle-averaged radial profiles for the 
H$\alpha$, IR, X-ray, and 6 cm (4835.1 MHz) radio total intensity emission 
in three sectors of the remnant.  In Figure \ref{irings2}, angle-averaged 
radial profile plots of the X-ray and steep- and flat-spectrum radio emission 
are shown for the same sectors.  There is a strong correspondence between 
these different emission components, but it varies from sector to sector.  
Some of the key patterns are as follows:  1.  The H$\alpha$ and IR have the 
same distribution (NNE, NW).  2.  The H$\alpha$ and IR are either in front of 
the X-ray and radio emission (NNE) or at the same leading edge (NW).  Note 
that the leading plateau of radio emission comes from parts of the western 
ear, not the bright shell.  3.  The X-ray emission corresponds in size, 
shape, and position to either one of the radio components (flat, NW; steep, 
S) or both (NNE).  4.  The flat-spectrum radio emission is either in front of 
(NW, S) or coincident with (NNE) the steep-spectrum radio emission.  For 
reference, circles 100$\arcsec$ in radius and centered at $\alpha_{2000} = 
17^{\mathrm{h}} 30^{\mathrm{m}} 41\fs25$ and $\delta_{2000} = -21\degr 
29\arcmin 29\farcs7$ are shown in Figure \ref{cxoi}.

In Figure \ref{azimuth}, we show azimuthal variations across the northern 
ring.  The H$\alpha$ and IR emission is concentrated to the northwest and 
northeast.  In contrast, the total and flat-spectrum radio emission is 
brightest directly north.  The X-ray emission is a hybrid between these two.  
The steep-spectrum radio emission exhibits similar azimuthal dependence to the 
H$\alpha$ and IR emission in the north but at a smaller radius (50$\arcsec - 
80\arcsec$).

\section{Discussion}
\label{sec:disc}

Our long-term goal is to understand the origins, evolution, and coupling 
between the multiple plasmas that exist in Kepler's SNR.  In the current 
work, we have used radio, X-ray, H$\alpha$, and IR images at a resolution of 
7$\farcs2$ to compare the signatures of these plasmas in the various 
wavelength regimes, along with the radio spectral index and polarization 
measurements.  At present, it is not possible to produce a single, coherent 
picture of the remnant's complex structures.  In this discussion, we identify 
when clear relationships exist among the multiwavelength structures, and when 
no single pattern emerges.  We explore how the multiple radiative signatures 
can be used as diagnostics for a variety of underlying physical conditions 
and processes in Kepler's SNR, as a step towards the long-term goal. 

One example of a clear pattern is the similarity of the H$\alpha$ and IR 
emission as shown in Figure \ref{cxoi}.  Although the H$\alpha$ emission 
results from mostly radiative processes in the northwest and mostly 
nonradiative processes elsewhere \citep{fbb89, blv91}, there is 12 $\micron$ 
thermal dust emission associated with both types of H$\alpha$ emission 
processes.  The position of the H$\alpha$ emission in front of the X-ray and 
radio emission is consistent with a circumstellar origin.  This is because 
the lifetime of a neutral atom that has passed through a shock front is short 
so any emission from the excitation of a neutral atom before it has been 
ionized by collisions with electrons in the postshock flow must occur 
immediately behind the shock \citep{ckr80}.  If the optical (including 
H$\alpha$) knots are circumstellar in origin then, \citet{dlb99, dlc01} 
conclude, based on the association of H$\alpha$ and IR emission, that the 
12 $\micron$ IR emission must also be circumstellar or interstellar in origin.

There are a number of factors which affect the local dynamics of the remnant 
and would prevent a single global set of relationships between the multiple 
observed wavelengths.  These include variations in external density over a 
variety of different scales, motion of the progenitor, and asymmetries in the 
explosion.  These will affect the relative strengths and location of the 
forward and reverse shocks, any bow shock features from the progenitor 
motion, and the details of the relativistic particle acceleration.  Instead 
of focusing on single, global patterns, we need to examine relationships in 
different locations of the remnant.

\begin{figure}[ht]
\epsscale{1}
\plotone{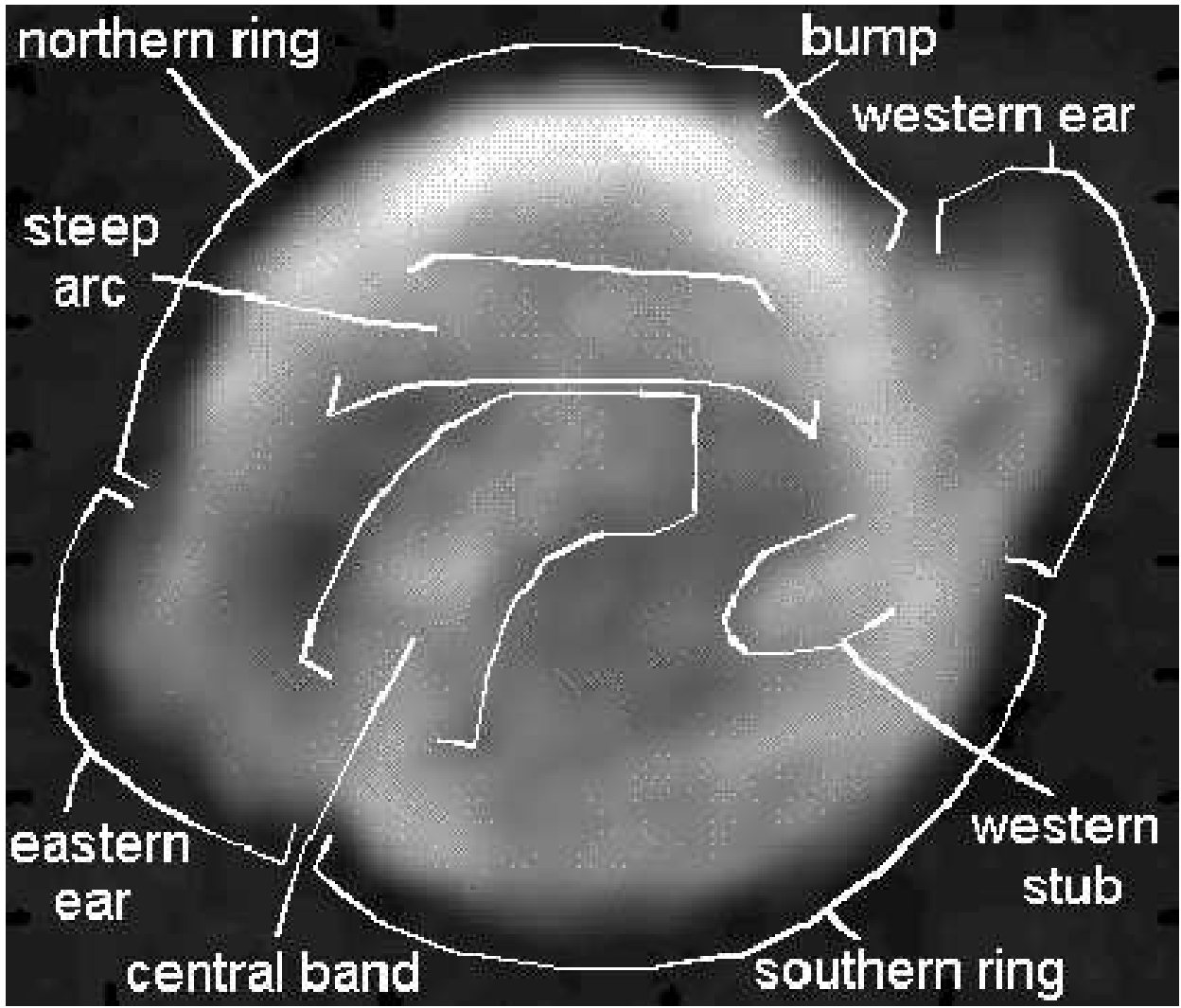}
\caption{Gray-scale image of Kepler's SNR at 6 cm.  The resolution 
is 7$\farcs$2.  The gray-scale brightness range is $-$0.35 to 35.78 mJy 
bm$^{-1}$.  \label{cband}}
\end{figure}

\begin{figure}[ht]
\epsscale{.60}
\plotone{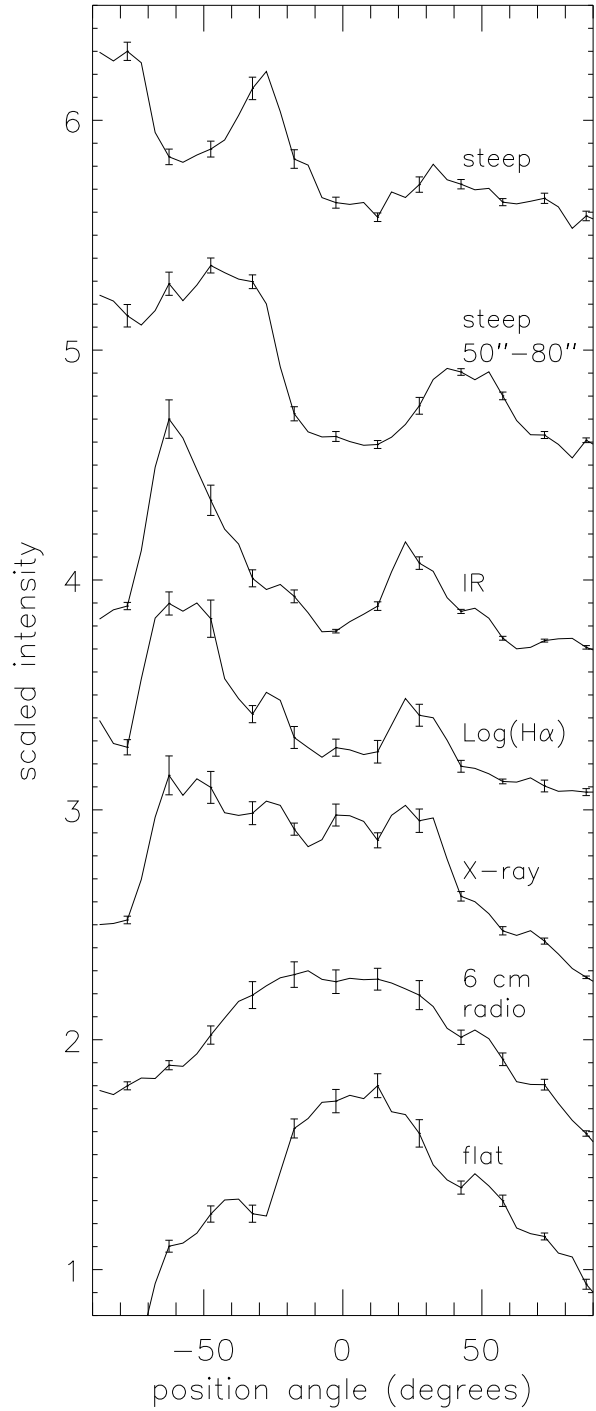}
\caption{Intensity averaged every 5$\degr$ in position angle for an annulus 
70$\arcsec - 100\arcsec$ in radius around the northern ring.  Note that the 
steep-spectrum radio emission has also been averaged in an annulus from 
50$\arcsec - 80\arcsec$.  The rms of the mean intensity is shown for every 
third independent azimuth bin.  The center used for this plot is the same 
as that in Figure \ref{irings}.
\label{azimuth}}
\end{figure}

We start by looking directly north where the radio emission is brightest as 
shown in the grey-scale images of Figures \ref{lband}, \ref{cband}, and 
\ref{cxoi} and the azimuthal profile of Figure \ref{azimuth}.  The two ways 
to enhance the synchrotron emission are to amplify the magnetic fields and 
to increase the number of synchrotron radiating electrons.  The radio 
emission, then, is a measure of the combination of magnetic field strength, 
thermal density (because the thermal electrons should provide the seed 
electrons which are accelerated by shocks to relativistic speeds), and the 
efficiency of the acceleration process.  However, Figure \ref{azimuth} shows 
that the thermal tracers, the H$\alpha$, IR, and X-ray emission, are all 
weaker directly north where the radio emission is brightest.  If the 
brightest radio regions are not areas of enhanced density, then either the 
magnetic fields must be stronger where the radio is brightest or the 
acceleration process is more efficient.

\begin{figure}[ht]
\epsscale{.60}
\plotone{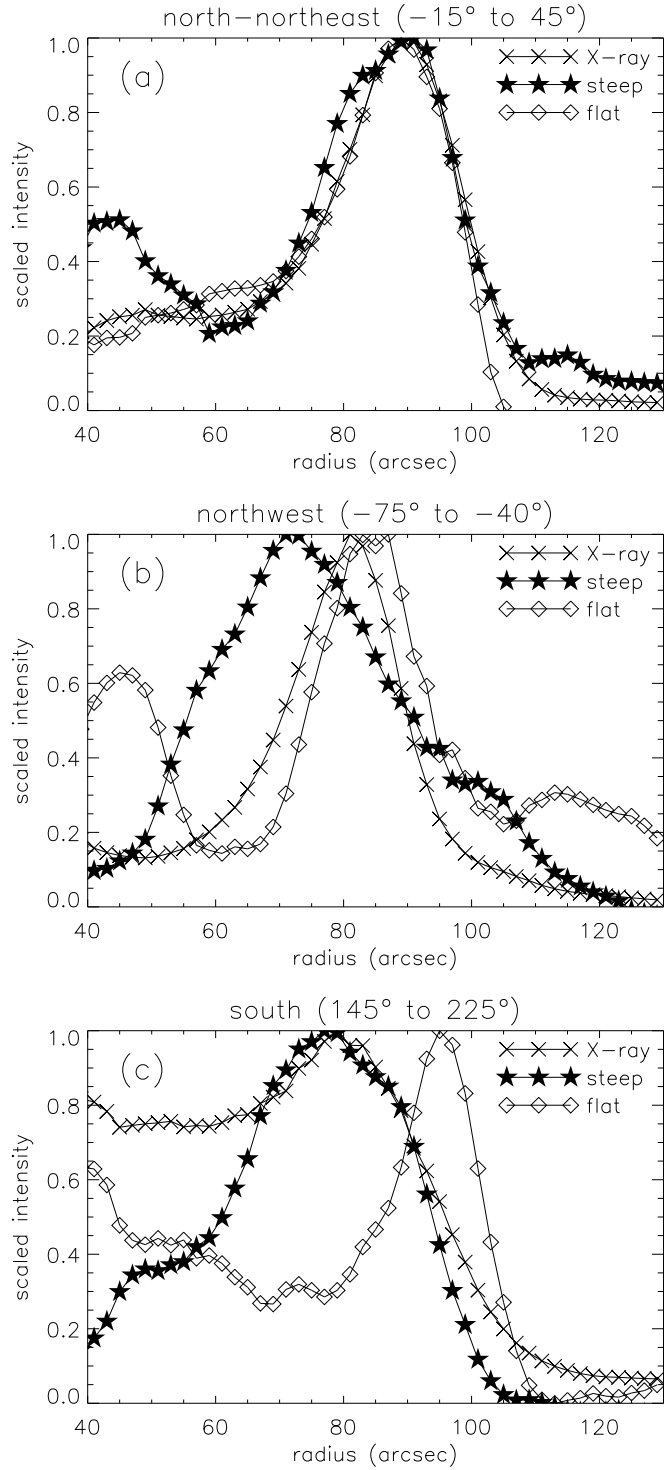}
\caption{The same as Figure \ref{irings} for X-rays, flat-spectrum radio, and 
steep-spectrum radio.
\vspace{4.5em}
\label{irings2}}
\end{figure}

Comparing the X-ray to the steep and flat spectral components of the radio 
emission reveals an interesting correlation.  As shown in Figure 
\ref{irings2}, in the southern portion of the ring, the X-ray clearly tracks 
the steep-spectrum radio emission.  In the north to northeast, the X-ray, 
flat-spectrum radio, and steep-spectrum radio have coincident leading edges.  
In the northwest, the X-ray more closely tracks the flat-spectrum radio 
emission.  In all cases, the flat-spectrum radio emission either has the same 
leading edge as or is in front of the steep-spectrum radio emission.  
Although the radial profiles do not allow us to cleanly separate the flat- and 
steep-spectrum radio emission, we suggest that the spectral components are 
superposed along the line of sight.  The alternative, that the regions 
actually represent an intermediate spectral index population, cannot be ruled 
out using spectral tomography.  Figure \ref{azimuth} shows that the X-ray, 
H$\alpha$, IR, and steep-spectrum radio emission have similar azimuthal 
profiles over $\approx$ 150$\degr$ in angle, and are therefore all likely 
responding to variations in the circumstellar density.

The difference in structure between the flat- and steep-spectrum radio 
emission suggests that we may be seeing a partial decoupling of the forward 
and reverse shocks.  Because the X-rays in the ROSAT image are mostly line 
emission from iron and silicon \citep{kt99}, they must be from shocked ejecta 
rather than shocked ISM.  The conditions (such as different temperatures and 
differing seed particle energy distribution) in the shocked ejecta may 
produce steeper synchrotron emission than the shocked ISM.  One problem here 
is that the northern ring should be more dense than the southern ring.  In 
this case, a reverse shock should have formed earlier in the north and there 
should be a greater separation between forward and reverse shocks there than 
in the south.  This is opposite to the observations shown in Figure 
\ref{irings2}.  However, if the forward shock has only very recently 
encountered denser ISM, its expansion would be slowed, allowing for less 
separation between the forward and reverse shocks.

\citet{ban87} proposed a model for the morphology of Kepler's SNR where the 
progenitor was a massive runaway star.  In this model, a bow shock forms in 
the direction of motion of the star (to the north).  The bow-shocked shell is 
reshocked by the forward shock and the north-south brightness asymmetry is a 
direct result of greater density in the north than in the south.  Numerical 
simulations \citep{bbs92} show that there is minimal separation between the 
bow-shocked shell and the reverse shock in the north.  To the south, the 
bow-shocked shell is less dense and thus less bright than in the north.  In 
this interpretation, the forward shock is not seen because it is now 
propagating into very low density ISM.  The material in the bow-shocked shell 
has a higher density and lower temperature than the stellar ejecta.  These 
conditions result in a higher Mach number (and thus a flatter spectral 
index) for a shock propagating through the bow-shocked material rather than 
the stellar ejecta.  We therefore, still associate the steep-spectrum radio 
emission with reverse-shocked material, however, the flat-spectrum radio 
emission is now associated with bow-shocked material.    

We now explore why the flat-spectrum radio component, which we suggest 
indicates the location of either the outer shock or the progenitor bow shock, 
does not show the same azimuthal structure as the thermal emission shows.  
Perhaps the anticorrelation seen in Figure \ref{azimuth} is due to a relative 
weakening of the particle acceleration at the forward shock when regions of 
high neutral density are encountered.  Ion-neutral friction in the upstream 
flow can damp Alfv\'{e}n wave growth \citep{ddk96}.  Alfv\'{e}n waves are 
necessary in the upstream flow to stop particles which have crossed the shock 
from escaping and reflect them back across the shock allowing the first order 
Fermi acceleration process to work.  Equation 46 from \citet{ddk96} sets the 
upper cutoff energy for ions and electrons to be reflected by Alfv\'{e}n 
waves:  
\begin{eqnarray*}
\frac{E}{1\;\mbox{TeV}} < 
\left(\frac{U}{10^3\;\mbox{km s$^{-1}$}}\right)^3 
\left(\frac{T}{10^4\;\mbox{K}}\right)^{-0.4}\\ \times
\left(\frac{n_n}{1\;\mbox{cm$^{-3}$}}\right)^{-1} 
\left(\frac{n_i}{1\;\mbox{cm$^{-3}$}}\right)^{0.5}
\left(\frac{\mathcal{P}_0}{0.1}\right).
\end{eqnarray*}
Here, $U$ is the shock velocity, $T$ is the preshock temperature, $n_n$ is 
the neutral number density, $n_i$ is the ion number density, and 
$\mathcal{P}_0$ is the dimensionless resonant cosmic ray pressure at the 
shock.  Any particles with energies higher than $E$ will escape upstream.  To 
determine the upper cutoff energy, we first take $n_n$ = 7 cm$^{-3}$ and $U$ 
=1500 km s$^{-1}$ from \citet{blv91}.  We then assume, for 600 pc above the 
galactic plane, that $T = 10^4$ K and $n_i$ = 1 cm$^{-3}$ (an ionization 
fraction of 13 \%).  Then, if the shock acceleration is efficient 
($\mathcal{P}_0 = 0.1$), the upper cutoff energy is 480 GeV.  

The above cutoff energy is too high to affect our radio measurements.  
Suppose, however, the shock acceleration is less efficient than assumed 
above, leading to a smaller $\mathcal{P}_0$.  A lower limit to the resonant 
cosmic ray pressure can be set by the presence of background cosmic rays 
which have an energy density of $\approx$ 1 eV cm$^{-3}$.  Dividing the 
background cosmic ray pressure by the ram pressure of upstream ions results 
in a dimensionless pressure of $4.7 \times 10^{-5}$.  Using this pressure, 
the upper energy cutoff now becomes 0.2 GeV.  The electrons radiating at 6 cm 
have energies of 4 GeV, if the magnetic field and particle energies are in 
equipartition.  Therefore, it is possible that Alfv\'{e}n wave damping is 
occurring to some degree in Kepler's SNR and this results in the 
anticorrelation of the flat-spectrum radio emission and the thermal 
emission.  It would be useful to explore the possible contribution of 
Alfv\'{e}n wave damping for spectral variations in other SNRs.

We now turn to the structural aspects of the remnant and the connection with 
underlying dynamics.  There are two primary classes of radio structure in the 
remnant as identified from the tomography images shown in Figure \ref{tomo} 
and labelled in Figure \ref{cband}.  The ring, consisting of the structures 
labelled as the northern ring and southern ring, is a composite of both flat 
and steep spectral components.  The non-ring features include the ears, 
central band, bump, steep arc, and western stub.  The non-ring features are 
all steep in spectral index which leads one to ask if they are spatially 
connected with each other and/or with the steep parts of the ring.  
Unfortunately, there does not seem to be an easy way to connect these 
features into a coherent structure.  This lack of connection is also evident 
in the rotation measure image of Figure \ref{rm} that does not show the same 
structures as the spectral index image.  We might expect spatially coherent 
structures to show a clear pattern in rotation measure, but none is apparent.

The ears, labelled in Figure \ref{cband}, may arise due to expansion into a 
lower density medium.   Pressure equilibrium in the presupernova ISM leads to 
higher temperatures in lower density areas.  Higher temperatures result in 
higher sound speeds and thus lower Mach numbers in these regions if the shock 
velocity is independent of density.  Lower Mach numbers indicate weaker shocks 
and thus steeper expected spectral indices.  To see if this hypothesis is 
plausible, consider the relation between spectral index and Mach number for 
first-order Fermi acceleration in the test particle limit, 
$\alpha=(M^2 + 3)/2(M^2 - 1)$.  If $\alpha=0.65(0.8)$, then $M=3.8(2.8)$.  The 
resulting density ratio between a flat spectral region in the north and a 
steep spectral region in the western ear is $n_f/n_s = (M_f/M_s)^2 = 1.8$ 
which is certainly within reason.  

The fact that the ears are 180$\degr$ apart indicates that, if the progenitor 
was really a massive star, the stellar wind may have been axisymmetric 
\citep{blc96}.  This model predicts the shapes of very young SNRs and might be 
extendable to Kepler's higher age.  However, if the progenitor was a runaway 
star, as suggested by \citet{ban87}, then the axisymmetry would be lost.  
There is evidence, however, that the progenitor's stellar wind has modified 
the circumstellar environment at the current shell radius because the optical 
knots are thought to be material ejected by that wind \citep{blv91}.  

An alternative source of the ear structures could be a strongly magnetized 
interstellar medium \citep{ir91,rt95}.  An SNR evolving in such a medium will 
become elongated along the direction of the field.  This may also explain the 
steeper spectral indices in the ears because shocks propagating parallel to 
the ambient magnetic field produce steeper spectra than shocks propagating 
perpendicular to the magnetic field \citep{fr90}.  However, this model cannot 
explain the almost spherical symmetry of the rest of the remnant and requires 
a magnetic field much stronger than is thought to exist in the ambient 
interstellar medium so far from the galactic plane.  Finally, the ears could 
also be due to collimated outflows from a central source as \citet{ggm98} 
suggest for G309.2-00.6.  However, no jet-like structure or compact object is 
identified in Kepler.  None of these models for the ears --- axisymmetric 
wind, interstellar magnetic fields, or collimated outflows --- explains the 
significant north-south brightness asymmetry seen in all bands.

The generally radial orientation of the magnetic field shown in Figure 
\ref{bfield} is consistent with what we expect from Rayleigh-Taylor (RT) 
instabilities in the decelerating remnant \citep{gul75, che76} and is common 
in young SNRs \citep{mil88}.  The deviations from the radial pattern are 
explained by \citet{dej89} as due to an inhomogeneous circumstellar medium 
(CSM).  \citet{dej89} used clump radii of $10^{17}$ cm which correspond to 
angular diameters of 2$\farcs$7 at the distance of Kepler which is smaller 
than our current resolution of 7$\farcs$2.  With a clumpy CSM, RT 
instabilities and MHD turbulence develop at every clump, which can account 
for both the observed magnetic field patterns and the brightness distribution 
in the remnant.  We might expect to see magnetic field patterns corresponding 
to brightness variations in the radio, X-ray, H$\alpha$, and IR images shown 
in Figure \ref{cxoi}.  No correlation is seen with radio brightness 
features.  No obvious correlation exists with the X-ray, H$\alpha$, and IR 
brightness either, but severe depolarization limits magnetic field 
measurements in the highest density regions.  If density variations (clumps) 
are responsible for non-radial magnetic field patterns, we might also expect 
to see a correspondence with spectral index as shown in Figure \ref{spix}, 
but we do not.    

\section{Conclusions}
\label{sec:conc}

In this paper we have presented a new epoch of radio spectral and polarization 
data for Kepler's SNR.  We have also compared our new radio data to H$\alpha$, 
X-ray, and IR images to better understanding the interactions between the 
nonthermal and thermal plasmas.  We identify the following primary results:

\begin{enumerate}

\item There is a general north-south asymmetry in all emission and 
polarization measures.

\item For the first time spatial spectral index variations are observed 
between 6 cm and 20 cm in Kepler's SNR.  The spectral indices range from 
$-$0.85 to $-$0.6 with a mean of $-$0.71.

\item The mean percent polarization is 3.5\% at 20 cm and 6\% at 6 cm.

\item Although there are interesting small-scale features such as the bump and 
the western ear, there is no clear large-scale pattern between spectrally 
steep features and rotation measure, depolarization, or thermal emission.

\item There are several strong correspondences between the radial and 
azimuthal profiles of the radio, X-ray, H$\alpha$, and IR emission in 
different parts of the remnant although there is no single, global pattern.

\item The striking similarity between the H$\alpha$ and IR images suggests 
that the 12 $\micron$ thermal dust emission must have the same origin as the 
optical emission, namely shock heating of the CSM/ISM.

\item The brightest thermal emission regions correspond to the strongest radio 
depolarization indicating that the thermal and nonthermal plasmas are mixed.

\item The flat-spectrum and steep-spectrum radio emission indicate forward- 
and reverse-shocked material, respectively, and indicate a partial decoupling 
of these shocks in the southern portion of the remnant.

\item The flat-spectrum radio emission may alternatively indicate bow-shocked 
material (reshocked by the forward shock) from the progenitor's motion 
through the ISM.

\item The anticorrelation in the azimuthal profiles of the flat-spectrum radio 
emission and the thermal emissions may indicate that Alfv\'{e}n wave damping 
is occurring to some degree in Kep\-ler's SNR.

\item The ears may result from expansion into a lower density medium and any 
model used to explain the ears must include axisymmetry.

\item The magnetic field is generally radial as expected, but small-scale 
magnetic field patterns do not correspond to brightness variations in the 
radio, X-ray, H$\alpha$, or IR.  There is also no correlation between magnetic 
field patterns and radio spectral index.

\end{enumerate}

To date there are still many unresolved issues about how the thermal and 
nonthermal plasmas are coupled in SNRs.  Many different correlations between 
measures exist, but often no clear pattern emerges.  Comparison with proper 
motion measurements is essential to help define why certain correlations exist 
in certain cases.  Future observations must also be conducted with 
\emph{Chandra} and SIRTF which will provide higher spatial resolution and 
allow us to more definitively determine the relationships between the thermal 
and nonthermal plasmas.

\acknowledgments

This work has been conducted with support by NSF grant AST 96-19438 at the 
University of Minnesota.  We wish to thank P.-O. Lagage and T. Douvion for 
providing an IR image , W. Blair for the H$\alpha$ image, and the ROSAT 
public archive for the X-ray data. We also wish to thank Udo Gieseler and Tom 
Jones for valuable conversations.  We acknowledge the use of NASA's 
\emph{SkyView} facility \footnote{\url{http://skyview.gsfc.nasa.gov}} located 
at NASA Goddard Space Flight Center for providing the DSS and 2MASS images 
which were used to register the H$\alpha$ and IR images.  We finally wish to 
thank the referee for many valuable comments.

\clearpage

\begin{deluxetable}{lcccc}
\tabletypesize{\scriptsize}
\tablecaption{VLA Observations \label{tbl1}}
\tablewidth{0pt}
\tablehead{
\colhead{} & 
\colhead{} & 
\colhead{} & 
\colhead{} & 
\colhead{On-Source Time} 
\\
\colhead{} & 
\colhead{} & 
\colhead{Wavelength} & 
\colhead{Bandwidth} & 
\colhead{per Frequency} 
\\
\colhead{Date} & 
\colhead{Configuration} & 
\colhead{(cm)} & 
\colhead{(MHz)} & 
\colhead{(minutes)}
}
\startdata
1997 May & B & 20 & 25\phn\phd & 101 \\
 & & \phn6 & & 362 \\
1997 Jul & C & 20 & 25\phn\phd & \phn43 \\
 & & \phn6 & & 145 \\ 
1997 Dec, 1998 Jan & D & 20 & 25\phn\phd & \phn15 \\
 & & \phn6 & & \phn77 \\
1998 Mar, Apr & A & 20 & 12.5 & \phn83 \\
\enddata
\end{deluxetable}

\clearpage

\begin{deluxetable}{ccc}
\tabletypesize{\scriptsize}
\tablecaption{Statistics for Radio Total Intensity Images \label{tbl2}}
\tablewidth{0pt}
\tablehead{
\colhead{} & 
\colhead{Integrated} & 
\colhead{Off Source} 
\\
\colhead{Frequency} & 
\colhead{Flux Density} & 
\colhead{RMS Noise}
\\
\colhead{(MHz)} &
\colhead{(Jy)} &
\colhead{(mJy bm$^{-1}$)}
}
\startdata
1285.0 & 15.4 & 0.27 \\
1365.1 & 14.8 & 0.34 \\
1402.5 & 14.6 & 0.39 \\
1464.9 & 14.2 & 0.29 \\
4835.1 & \phn6.1 & 0.12 \\
4885.1 & \phn6.1 & 0.15 \\
\enddata
\end{deluxetable}

\clearpage

\clearpage

\clearpage

\clearpage

\clearpage

\clearpage

\clearpage

\clearpage

\clearpage

\clearpage

\clearpage

\clearpage

\clearpage

\clearpage

\clearpage

\clearpage

\clearpage 

\clearpage

\end{document}